\documentclass[]{aastex63}

\newcommand{\beq}{\begin{equation}}
\newcommand{\eeq}{\end{equation}}
\newcommand{\beqn}{\begin{eqnarray}}
\newcommand{\eeqn}{\end{eqnarray}}

\newcommand{\WSU}{\affiliation{Department of Physics \& Astronomy,
	Washington State University, Pullman, Washington 99164, USA}}
\newcommand{\UNH}{\affiliation{Department of Physics \& Astronomy, University of New Hampshire, 9 Library Way, Durham NH 03824, USA}}
\newcommand{\TAPIR}{\affiliation{TAPIR, Walter Burke Institute for Theoretical Physics, MC 350-17, California Institute of Technology, Pasadena, California 91125, USA}}
\newcommand{\Cornell}{\affiliation{Cornell Center for Astrophysics and Planetary Science, Cornell University, Ithaca, New York, 14853, USA}}
\newcommand{\MPI}{\affiliation{Max-Planck-Institut fur Gravitationsphysik, Albert-Einstein-Institut, D-14476 Potsdam, Germany}}

\shorttitle{MC transort in SpEC}
\shortauthors{Foucart et al.}

\begin{document}
\title{Implementation of Monte-Carlo transport in the general relativistic SpEC code}

\correspondingauthor{Francois Foucart}
\email{francois.foucart@unh.edu}

\author{Francois Foucart} \UNH
\author{Matthew D. Duez} \WSU
\author{Francois H\'{e}bert} \TAPIR
\author{Lawrence E. Kidder} \Cornell
\author{Phillip Kovarik}\UNH
\author{Harald P. Pfeiffer} \MPI
\author{Mark A. Scheel} \TAPIR

\nocollaboration{7}

\begin{abstract}
Neutrino transport and neutrino-matter interactions are known to play an important role in the evolution of neutron star mergers, and of their post-merger remnants. Neutrinos cool remnants, drive post-merger winds, and deposit energy in the low-density polar regions where relativistic jets may eventually form. Neutrinos also modify the composition of the ejected material, impacting the outcome of nucleosynthesis in merger outflows and the properties of the optical/infrared transients that they power (kilonovae). So far, merger simulations have largely relied on approximate treatments of the neutrinos (leakage, moments) that simplify the equations of radiation transport in a way that makes simulations more affordable, but also introduces unquantifiable errors in the results. To improve on these methods, we recently published a first simulation of neutron star mergers using a low-cost Monte-Carlo algorithm for neutrino radiation transport. Our transport code limits costs in optically thick regions by placing a hard ceiling on the value of the absorption opacity of the fluid, yet all approximations made within the code are designed to vanish in the limit of infinite numerical resolution. We provide here an in-depth description of this algorithm, of its implementation in the SpEC merger code, and of the expected impact of our approximations in optically thick regions. We argue that the latter is a subdominant source of error at the accuracy reached by current simulations, and for the interactions currently included in our code. We also provide tests of the most important features of this code. 
\end{abstract}

\section{Introduction}

The joint detection of gravitational waves and electromagnetic signals from the first confirmed neutron star merger observation, GW170817~\citep{2017ApJ...848L..13A,2017Sci...358.1559K,2017ApJ...848L..19C,2017Natur.551...75S,2017ApJ...848L..16S,Cowperthwaite:2017dyu}, recently demonstrated the potential power of these systems to study general relativity, nuclear physics, astrophysical nucleosynthesis, and the properties of compact objects. However, theoretical uncertainties in the amount of mass ejected by a given merger~\citep{Kruger:2020gig} and its composition~\citep{Wanajo2014,Foucart:2018gis}, radiation transport in that ejecta~\citep{Heinzel:2020qlt}, the outcome of nucleosynthesis~\citep{2013ApJ...775...18B}, and the energy that released by nuclear reactions in the ejecta~\citep{Barnes:2016} limit the amount of information that we can extract from merger observations.

Numerical simulations of neutron star mergers play an important role in our ability to analyze these systems. Ideally, we would like simulations that predict, for any binary merger, the amount of matter ejected, as well as the composition, velocity, and geometry of the outflows. Indeed, these are the main determinant of the outcome of r-process nucleosynthesis in the outflows~\citep{Lippuner2015} and of the brightness, time evolution, and color of kilonovae~\citep{2013ApJ...775...18B}. While numerical simulations of neutron star mergers have made a lot of progress over the last two decades (see e.g.~\cite{baiotti2016,Shibata:2019wef,Dietrich:2020eud,Ciolfi:2020huo} for reviews), three important problems continue to limit our ability to reliably predict the properties of matter outflows: our inability to resolve magnetic fields~\citep{Kiuchi2014}, our approximate treatment of neutrino transport~\citep{Foucart:2018gis}, and a lack of consistency between merger and post-merger simulations that makes it difficult to interpret the result of the longest 3D post-merger simulations currently at our disposal~\citep{Siegel:2017nub,Fernandez:2018kax,Christie:2019lim}. We will focus here on the issue of neutrino transport.

Neutrinos play a number of important role in the evolution of neutron star mergers, and particularly of their post-merger remnants. First, neutrinos are the main source of cooling of post-merger remnants, with neutrino luminosities peaking at $L_{\nu}\sim 10^{53-54}\,{\rm erg/s}$ and remaining at these levels for $\sim (10-100)\,{\rm ms}$~\citep{Sekiguchi:2011zd,Foucart:2016rxm,Fujibayashi:2020qda}. Neutrino cooling plays a critical role in setting the thermodynamical properties of post-merger accretion disks, and in particular in limiting their typical thickness to $H/R\sim 0.2-0.3$ during the first $\sim 100\,{\rm ms}$ of post-merger evolution~\citep{Fernandez:2020oow}. Second, emission and absorption of electron neutrinos and antineutrinos ($\nu_e$ and $\bar \nu_e$) modifies the relative number of protons and neutrinos in the post-merger remnant and matter outflows. This is typically parametrized by the electron fraction $Y_e=n_p/(n_n+n_p)$, with $n_{p,n}$ the number density of protons and neutrons. The electron fraction is a crucial parameter in determining the outcome of r-process nucleosynthesis in the outflows and the color/duration of kilonovae~\citep{2013ApJ...775...18B,Lippuner2015}, and $Y_e$ tends to be strongly underestimated by approximate transport scheme that do not properly account for neutrino absorption~\citep{Wanajo2014,Foucart:2018gis}. Finally, neutrino-antineutrino pair annihilation in the low-density polar regions can deposit a significant amount of energy above the remnant~\citep{Just:2015dba,Fujibayashi:2017abc}, and may contribute to the formation of a baryon-free zone in that region and the eventual production of a relativistic jet.

The inclusion of neutrino-matter interactions in merger simulations remain a relatively recent event. A first leakage scheme~\citep{2010CQGra..27k4107S}, inspired from methods developed for Newtonian disk simulations and supernovae~\citep{1997A&A...319..122R,Rosswog:2003rv}, was implemented about a decade ago. Leakage algorithms provide order-of-magnitude estimates of the energy and lepton number leaving from a given point, but do not easily account for transport of neutrinos from one point to another and neutrino absorption (although more advanced leakage schemes have been developed to approximately take these effects into account in Newtonian simulations~\citep{Perego:2016}). The total neutrino luminosity can be captured within factors of a few by a leakage scheme~\citep{FoucartM1:2016}, but the composition of the outflows cannot be reliably measured in leakage simulations~\citep{Wanajo2014,FoucartM1:2016}. Leakage schemes however have the distinct advantage of being inexpensive to use in simulations, and remain a common way to approximately treat neutrino-matter interactions~\citep{Deaton2013,Neilsen:2014hha,Cipolletta:2020kgq}.

Moment schemes are the most common way to approximately include neutrino transport in merger simulations, going beyond the order-of-magnitude cooling captured by leakage schemes. In a moment algorithm, we evolved moments of the distribution function of neutrinos (taken in momentum space), e.g. the energy density and momentum density of neutrinos~\citep{thorne80,shibata:11}. Approximate analytical expressions are then used to close the transport equations, providing higher order moments and, for simulations using an energy integrated moment scheme, an estimate of the neutrino energy spectrum. Moment simulations used in merger simulations have so far used energy-integrated moments~\citep{Sekiguchi:2015,FoucartM1:2015,Radice:2016}. Moment schemes with an energy discretization have been used in supernova simulations~\citep{roberts:16b}, but may be difficult to use and/or lack accuracy in systems with rapid changes in the velocity of the background fluid. Moment schemes have the advantage to be relatively simple to implement in relativistic hydrodynamics simulations, because the form of the evolution equations is very similar to what is used to evolve the fluid variables. The cost of evolving neutrinos is comparable to the cost of evolving the fluid, and moment schemes are expected to be very accurate in optically thick regions. Their known disadvantages include the strong dependence of the evolution of $Y_e$ on the chosen neutrino energy spectrum~\citep{Foucart:2016rxm}, the creation of unphysical shocks in regions where neutrino beams cross (typically in the polar regions,~\cite{Foucart:2018gis}), and the approximation required when computing reaction rates that depend on the direction of propagation of neutrinos (e.g. pair annihilation;~\cite{Fujibayashi:2017abc,Foucart:2018gis}). More importantly, because moment schemes do not converge to the correct solution to the equations of radiation transport, we cannot reliably estimate errors in simulations without comparing them to more advanced radiation transport schemes. Moment schemes {\it may} be sufficiently accurate for many of our current needs, but we cannot verify whether this is the case without going further in our modeling of neutrinos.

Going to an actual evolution of Boltzmann's equations of radiation transport is generally a more expensive proposition. Boltzmann's equation requires the evolution in time of a 6-dimensional distribution function for each species of neutrinos, with stiff source terms that couple neutrinos to the fluid, as well as couplings between neutrinos of different energy, direction of propagation, and/or species. A brute force discretization of this equation on a 6D finite difference grid is unlikely to be affordable any time soon. A few alternative methods have been proposed for neutron star mergers, including expansion of the momentum space distribution onto spherical harmonics~\citep{radice:13}, lattice-Boltzmann methods for radiation transport~\citep{Weih:2020qyh}, using Monte-Carlo transport to close the moment equations~\citep{Foucart:2017mbt}, and full Monte-Carlo transport of neutrinos~\citep{Foucart:2020qjb}. So far, only the latter has been directly used in a merger simulation, with comparison between Monte-Carlo and moment evolutions showing differences at the $10\%$ level in most observables (\cite{Foucart:2020qjb}, in simulations that ignored pair annihilation processes).

In this manuscript, we provide a detailed description of the Monte-Carlo radiation transport algorithm implemented in the SpEC merger code\footnote{http://www.black-holes.org/SpEC.html}, which we used in neutron star merger simulations in~\cite{Foucart:2020qjb}. We discuss in particular the approximations used to circumvent known issues with the use of Monte-Carlo algorithms in optically thick regions, which are the main difficulty encountered when attempting to use Monte-Carlo methods in neutron star mergers. Our algorithm is meant first and foremost to allow for affordable Monte-Carlo transport while retaining acceptable discretization and sampling errors. We provide here discussions of the trade-offs that this implies. In addition to the methods used in~\cite{Foucart:2020qjb}, we also discuss a simple methods to account for neutrino pair annihilation in low-density regions in our Monte-Carlo simulations, as well as an important change in the choice of time step used for neutrino propagation that reduces discretization errors in regions of high scattering opacity.  Sec.~\ref{sec:methods} discusses our numerical algorithm, while Sec.~\ref{sec:tests} presents important tests of our methods. Table~\ref{tab:symbols}, at the end of this manuscript, summarizes the symbols used in multiple sections of this document.

\section{Numerical Methods}
\label{sec:methods}

\subsection{Distribution Function}
\label{sec:fnu}

When evolving the equations of radiation transport in general relativistic simulations, we aim to determine the distribution function of particles $f(t,x^i,p^\mu)$. If we treat each particle {\it k} as a well-localized point particle with position $x^i_k(t)$ and 4-momentum $p_\mu^k(t)$, the distribution function is
\beq
f(t,x^i,p^\mu) = \sum_k \delta^3\left(x^i-x^i_k(t)\right)\delta^3\left(p_i-p_i^k(t)\right)
\eeq
where the sum is over all particles in our 4-dimensional spacetime, and $p_i^k$ are the spatial components of the 4-momentum of particle {\it k}. We note that while neither of the Dirac distributions is covariant, their product is (see e.g.~\cite{Ryan2015}). The distribution function is thus a well defined scalar distribution.\footnote{This treats neutrinos as classical point particles, which of course is not correct; but this will be sufficient for our simulations.}

Practically, there are far too many particles to evolve each of them individually. Most methods used to evolve $f$ thus smooth out the distribution function over a volume containing a large number of particles, and then discretize the distribution function using, e.g., finite difference or spectral methods. The distribution function then follows Boltzmann's equation of radiation transport
\beq
p^\alpha \left[ \frac{\partial f}{\partial x^\alpha}-\Gamma^\beta_{\alpha\gamma} p^\gamma \frac{\partial f}{\partial x^\beta}\right] = \left[ \frac{df}{d\tau}\right]_{\rm collisions},
\eeq
with $\Gamma^\alpha_{\beta\gamma}$ the Christoffel symbols of our spacetime. In this equation, the left-hand side indicates that free-streaming particles propagate along geodesics. The right-hand side includes all collision terms. Boltzmann's equation thus requires us to evolve a 6-dimensional function in time. Additionally, the right-hand side include terms coupling particles with different momenta (scattering events) as well as potentially stiff coupling terms between the particles and a hot/dense fluid (emission/absorption). In the case of neutrinos, we also have a separate distribution function for each species of neutrinos and antineutrinos ($\nu_e, \bar \nu_e,\nu_\mu, \bar\nu_\mu, \nu_\tau, \bar\nu_\tau$), and these distribution functions may themselves be coupled through collision terms (e.g. pair annihilation). In a high-dimensional space, and for problems lacking obvious symmetries, this quickly becomes very expensive\footnote{Including neutrino oscillations would be even costlier, and has only been done so far in post-processing, e.g. in ~\cite{2012PhRvD..86h5015M,2016PhLB..752...89W}. Neutrino oscillations could nevertheless play a role in setting the composition of some merger outflows.}. 

Monte-Carlo methods take a different approach to this problem. In a Monte-Carlo algorithm, we create $n_p$ ``superparticles'' (hereafter {\it packets}) that each represent a large number of particles. Each packet has a single position and momentum. We then approximate the distribution function as
\beq
f(t,x^i,p^\mu) = \sum_{k=0}^{n_p} N_k \delta^3\left(x^i-x^i_k(t)\right)\delta^3\left(p_i-p_i^k(t)\right)
\eeq
with $N_k$ the number of particles represented by packet {\it k}, and $x^i_k(t), p_{\mu}^k(t)$ its assumed position and 4-momentum. The packets aim to provide an unbiased sample of the underlying distribution function. Monte-Carlo algorithms have the advantage of being very adaptive: if most particles are in a small region of phase space, then most packets will also be in that region of phase space. At low resolution (i.e. for a small number of packets), computational resources are used very efficiently. The evolution of Monte-Carlo packets is also fairly intuitive: packets are emitted, move along geodesics, scatter, and get absorbed as individual particles, with probabilities chosen so that the packets remain, as much as possible, an unbiased sample of the distribution function. 

Monte-Carlo methods also come with important drawbacks: simulations become non-deterministic;  the distribution function at a given point of phase space is only known up to the sampling noise of the method; that sampling noise converges away very slowly with increased computational resources (as $n_p^{-1/2}$); and the method can quickly become expensive (and potentially unstable) in regions where the mean free path of the particles is very small compared to the scale of the system being studied. This last issue is what makes Monte-Carlo methods difficult to implement for neutrino transport in merger simulations: the hot, dense regions formed during the merger of two neutron star cannot be evolved without significant modifications to standard Monte-Carlo methods.  

In the following sections, we provide an in-depth description of the general relativistic Monte-Carlo algorithm for neutrino transport implemented in the SpEC merger code, and discuss the strategies used in that code to mitigate the cost of evolving regions where neutrinos are strongly coupled to the fluid. We note that at this point, our algorithm is mainly designed to keep the cost of simulations manageable. Accordingly, {\it many of the choices made in the development of this algorithm are optimized for simulations using a small number of Monte-Carlo packets}. We will attempt to point out where better choices could be made in the future as more computational resources become available. Nevertheless, the methods are designed so that increased computational resources allow us to solve Boltzmann's equation more and more accurately. This is an important distinction with respect to approximate transport schemes such at the two-moment formalism that has so far been the state of the art for neutrino transport in merger simulations~\citep{Wanajo2014,Sekiguchi:2016,FoucartM1:2015,Foucart:2016rxm} : even with infinite computational resources, the two-moment formalism would not converge to a solution of Boltzmann's equation. 

\subsection{Stress-energy tensor and moments}
\label{sec:mom}

In our Monte-Carlo algorithm, we will often need to calculate the stress-energy tensor of neutrinos, as well as various moments of their distribution function. The general relativistic stress energy tensor is, in the Monte-Carlo formalism~\citep{Ryan2015}
\beq
T_{\mu\nu}(t,x^i) = \sum_{k=1}^{n_p} N_k \frac{p_\mu^k p_\nu^k}{\sqrt{-g}p^t_k} \delta^3\left(x^i-x^i_k(t)\right)
\eeq
with $g$ the determinant of the spacetime metric. An observer with 4-velocity $u^\mu$ measure the corresponding energy density
\beq
J = T_{\mu\nu}u^\mu u^\nu = \sum_{k=1}^{n_p} N_k \frac{\nu_k^2}{\sqrt{-g}p^t_k} \delta^3\left(x^i-x^i_k(t)\right)
\eeq
with $\nu_k = -p_\mu^k u^\mu$ the energy of neutrinos in packet $k$ as measured by our observer. The average energy density within a region of coordinate volume $V$ (e.g. a grid cell) is then
\beq
J = \sum_{k \in V} N_k \frac{\nu_k^2}{\sqrt{-g}V p^t_k} 
\eeq
with the sum now including only packets located within the volume $V$. Similarly, the average linear momentum $H_\alpha = -T^{\mu\nu} u_\mu (g_{\nu \alpha}+u_\nu u_\alpha)$ measured by an observer with 4-velocity $u^\mu$ is
\beq
H_\alpha =  \sum_{k \in V} N_k \frac{\nu_k \left(p_\alpha^k - \nu_k u_\alpha\right)}{\sqrt{-g}V p^t_k}. 
\eeq
By construction, $H_\alpha u^\alpha=0$. To couple particles with the fluid, we will also need to compute terms of the form
\beq
\int dt \kappa J 
\eeq
with $\kappa$ an opacity that depends on the position and momentum of a particle. We get
\beq
\int dt \kappa J  = \sum_{k=1}^{n_p} \kappa_k N_k \frac{\nu_k^2}{\sqrt{-g}p^t_k} \Delta t_k \delta^3\left(x^i-x^i_k(t)\right)
\eeq
with $\kappa_k$ the opacity experienced by packet $k$, which we assume to be constant during a time step in our algorithm, and $\Delta t_k$ the time interval within the integration domain during which packet $k$ existed. If we now compute the average value of this integral within a grid cell of volume $V$, we get
\beq
\int dt \kappa J  = \sum_{k \in V} \kappa_k N_k \frac{\nu_k^2}{\sqrt{-g}Vp^t_k} \Delta t_k.
\eeq
Typically, we will calculate these quantities either in the fluid frame (where $u^\mu$ is the fluid velocity) or in the simulation frame ($u^\mu=n^\mu$, with $n^\mu$ the unit normal to a constant-$t$ slice).

\subsection{Overview of the algorithm}

The Monte-Carlo algorithm implemented in the SpEC code evolves the equations of radiation transport coupled to Einstein's equation of General Relativity and to the general relativistic equations of (magneto)hydrodynamics. The methods used to evolve the metric and fluid variables are described in more detail in~\cite{Duez:2008rb,Foucart:2013a}. The metric is evolved using pseudospectral methods, and the fluid using high-order shock capturing methods on a separate finite volume grid with fixed mesh refinement (nested cubes, with each level of refinement decreasing the grid spacing by a factor of 2). The main improvement made to our code since the publication of~\cite{Foucart:2013a} is that we now allow the metric and fluid evolution to use different time steps, and different time stepping methods. In merger simulations that include neutrino transport, the metric is generally evolved using a third-order Runge-Kutta algorithm with adaptive time stepping, while the fluid uses a second-order Runge-Kutta algorithm with fixed Courant factor; these choices can however be modified at run time. For example, higher-order methods for the evolution of the fluid are used in simulations with higher accuracy requirements and less microphysics, e.g. the simulations used to test and calibrate gravitational wave models~\citep{Foucart:2018inp}. The time step may also be reduced to ensure stability of the control system used to keep the center of the compact objects fixed on our grid, and/or to keep the excised region around a black hole singularity as a sphere of constant radius in grid coordinates~\citep{Hemberger:2012jz}. The time step on the pseudospectral grid is always smaller or equal to the time step on the finite volume grid, and each finite volume grid step corresponds to an integer number of pseudospectral time steps. The metric terms needed for the evolution of the fluid and the fluid variables needed to construct the stress-energy tensor are communicated between our two grids at the end of each finite volume time step, and linearly extrapolated in time using the last two communicated values when values at different times need to be estimated. 

As neutrinos exchange momentum with the fluid, their evolution is most tightly linked to the fluid grid. In our code, neutrino packets are assigned to a specific cell of the finite volume grid, and all neutrinos in a given cell are evolved on the same processor. However, {\bf that processor is not necessarily the same as the one evolving the fluid variables}. Initially, the Monte-Carlo packets are evolved on the same processor as the fluid cell they live in, but as the evolution proceeds some cells may be ``loaned'' to other processors to improve load-balancing. This is crucial to maintain good performance of the code: in a typical simulations, most of the Monte-Carlo packets are located close to the hottest regions of the fluid, and keeping the packets tied to the processor evolving the corresponding fluid cell quickly leads to very poor load-balancing and wasted computational resources. 

The Monte-Carlo algorithm is called at the end of each time step taken on the finite volume grid, and uses a split time step algorithm. Schematically, the algorithm proceeds as follow:
\begin{enumerate}
\item Check the time difference $\Delta t$ between the current time and the end of the last Monte-Carlo step. If $\Delta t < C \Delta t_{c,\rm min}$, with $\Delta t_{c,\rm min}$ the shortest light-crossing time of the cells on the finite volume grid, the algorithm does nothing. Otherwise, we proceed to Step 2. We typically choose $C=0.5$.
\item Zero all variables used to keep track of momentum transfers between the neutrinos and the fluid, and send the fluid and metric variables needed for the evolution of the neutrinos from the processors owning the fluid data to the processor responsible for the evolution of neutrinos. All communications are performed using asynchronous MPI calls that only involve those two processors. The algorithm can proceed to Step 3 as soon as all MPI send and receive requests on a processor have been posted, but before the data expected from all other processors is actually received.
\item Compute the emission rate of neutrinos and the absorption / scattering opacities in all cells that have {\it not} been borrowed from another processor. The current algorithm models neutrino-matter interactions through an emissivity, an absorption opacity, and an elastic scattering opacity for each neutrino species. Accordingly, it cannot yet take into account reactions that do not fit in these categories (e.g. inelastic scattering). Neutrino-antineutrino pair annihilations in low-density regions are treated separately, as discussed below.
\item Emit packets in cells that have not been loaned to another processor, according to the emissivity computed in Step 3. Track the corresponding momentum exchanges between the fluid and neutrino sectors.
\item Evolve packets that started the time step in a cell that has not been loaned to another processor, or that were just emitted in such a cell. This includes propagating packets along geodesics, as well as the absorption and scattering of packets through interactions with the fluid. Track the corresponding momentum exchanges between the fluid and neutrino sectors, as well as any moment of the neutrino distribution function needed for other calculations.
\item Wait for the data sent during Step 2 to be received on the current processor, then repeat Steps 3-5 for cells loaned by another processor to the current processor.
\item Communicate all information about momentum exchanges and moments of the neutrino distribution function between processors.
\item If desired, perform neutrino-antineutrino pair annihilations and communicate information about momentum exchanges back to the processor owning the fluid data 
\item Check load-balancing, and improve the distribution of grid cells between processors if needed.
\item Update the fluid variables, accounting for momentum exchanges between the fluid and neutrino sectors.
\end{enumerate}
We provide more detail on each of these steps in the following sections. Some of the methods used here were already presented when we used an earlier version of our Monte-Carlo algorithm to close the evolution equations of an approximate two-moment scheme~\citep{Foucart:2017mbt}; we repeat them here for completeness.

\subsection{Propagation of packets}

To propagate Monte-Carlo packets, we need to evolve the position $x^i$ and 4-momentum $p^\mu$ of packets along geodesics. In our code, we neglect the mass of neutrinos and evolve packets along null geodesics. We work in the 3+1 formalism, where 
\beq
ds^2 = g_{\mu\nu} dx^\mu dx^\nu = -\alpha^2 dt^2 + \gamma_{ij} (dx^i + \beta^i dt)(dx^j + \beta^j dt),
\eeq
$g_{\mu\nu}$ is the spacetime metric, $\gamma_{ij}$ the spatial metric on constant-$t$ slices, $\alpha$ the lapse scalar, and $\beta^\mu = (0,\beta^i)$ the shift vector.
A convenient form for the evolution equations is~\citep{Hughes1994}:
\beqn
\frac{dx^i}{dt} &=& \frac{p^i}{p^t} = \gamma^{ij} \frac{p_j}{p^t} - \beta^i\\
\frac{dp_i}{dt} &=& -\alpha p^t \partial_i \alpha + p_j \partial_i \beta^j - \frac{1}{2} p_j p_k \partial_i\gamma^{jk}.
\eeqn
In this formalism, we evolve $x^i$ and $p_i$, and recompute $p^t$ as needed using the fact that, for a null vector
\beq
p^t = \frac{\sqrt{\gamma^{ij}p_i p_j}}{\alpha}.
\eeq
In our code, any given packet knows its current time $t_k$, its number of neutrinos $N_k$, the species of these neutrinos, and the cell on the fluid grid within which it started the current time step. It also knows its evolved position $x^i$ and spatial momentum $p_i$. The metric and metric derivatives are assumed constant within a finite volume cell. It would certainly be possible to improve on this and interpolate the value of the metric at the actual position of the packet with higher-order methods, yet this would require performing that interpolation at every intermediate step of the evolution of a packet, for every packet. This is a significant cost increase, when compared to the use of the known values of the metric at the center of each cell. We will see in our tests that the resulting error in the propagation of packets is small, indicating that at the current accuracy of our Monte-Carlo algorithm our low-order methods are sufficient. 

When evolving these equations in time, we use a second-order Runge-Kutta algorithm. The time step used for this evolution is the smallest of
\begin{itemize}
\item the time $\Delta t_{\rm step}$ needed to reach the end of the desired Monte-Carlo step
\item the times $\Delta t_{\rm a,s}$ to the next absorption / scattering event (see Sec.~\ref{sec:coll})
\item the time $\Delta t_{\rm cell} = \tilde f \Delta t_c$, with $\Delta t_c$ the minimum light-crossing time of the current grid cell, and $\tilde f$ a parameter chosen to avoid going too far out of a cell boundary during a time step. We set $\tilde f= f_{\rm grid} + f_{\rm min}$, with $f_{\rm grid}$ the distance between the packet location and the cell boundary in units of the grid spacing, $f_{\rm min}=\max{(0.03,\frac{0.1}{\tau_{\rm max}})}$, and $\tau_{\rm max}$ the maximum optical depth of a cell, considering only the current cell and its immediate neighbors. The optical depth used here includes both the scattering and absorption optical depth. We see that this stops a packet from moving too far out of its current cell when in an optically thick region, while we let packets propagate for the full time step in optically thin regions.
This condition is mainly aimed at improving accuracy in regions where the optical depth varies significantly between neighboring cells, which is why we ignore it in optically thin regions.
\end{itemize}
After taking this time step, we either move on to the next packet (if using $\Delta t_{\rm step}$ or $\Delta t_a$), perform a scattering and continue the evolution ($\Delta t_s$), or continue to propagate the packet after possibly moving it to a different grid cell ($\Delta t_{\rm cell}$). 

We note that in our first merger simulation using this Monte-Carlo code~\citep{Foucart:2020qjb}, we only considered $\Delta t_{\rm step}$ and $\Delta t_{\rm a,s}$, and ignored $\Delta t_{\rm cell}$. This led to cheaper evolutions and a more streamlined algorithm, at the cost of decreased accuracy in regions of high scattering opacities (see optically thick sphere and spherical collapse tests). 
Allowing packets to change cells in the middle of a Monte-Carlo time step improves the accuracy of the code, but at a cost. Now, a packet may have to be moved from its original grid cell to a neighboring cell during packet propagation. This means that we need to include one layer of ``ghost zone'' cells in the Monte-Carlo algorithm. In particular, when loaning a grid cell to another processor, we need to make sure that the new processor has access to the fluid and metric variable for that cell and for all neighboring cells. Additionally, the Monte-Carlo algorithm may deposit momentum in a ghost zone, and that information has to be communicated back to the processor owning the corresponding live cell. This makes parallelization of the code more difficult. Nevertheless, the significant improvement in the accuracy of the evolution of heavy-lepton neutrinos observed in test problems with that updated method motivates its use in future simulations.

Rather than including all neighboring cells in ghost zones (including cells that only share an edge or vertex with the current cell), we can also limit ourselves to the 6 neighbors sharing a face with the current cell. This seems to provide nearly the same benefits at a smaller communication cost. A packet moving in a neighboring cell that does not share a face with its original cell can then be randomly assigned to one of the closest available cells, using the metric/fluid variables of that cell. The difference between these methods would presumably become more noticeable at higher resolution.

\subsection{Collision terms}
\label{sec:coll}

So far, we have only considered the purely deterministic evolution of packets along geodesic. The probabilistic nature of the Monte-Carlo algorithm comes in the creation of packets and in their interactions with the fluid particles and other neutrinos, i.e. the collision terms in Boltzmann's equation. In this section, we focus on the basic treatment of these terms in our Monte-Carlo algorithm, ignoring the complications that arise in high opacity regions. We discuss the treatment of high-absorption and high-scattering opacity regions in Sec.~\ref{sec:highKs}-\ref{sec:highKa}.

\subsubsection{Tabulated reaction rates}

In the simulations presented in this work, we use the NuLib library~\citep{OConnor2010} to generate tabulated values of the emissivity $\eta$, absorption opacity $\kappa_a$, and elastic scattering opacity $\kappa_s$ experienced by Monte-Carlo packets. We take into account the charged current reactions
\beqn
p + e^- &\leftrightarrow& n + \nu_e\\
n + e^+ &\leftrightarrow& p + \bar \nu_e
\eeqn
as well as elastic scattering of all types of (anti)neutrinos on neutrons, protons, alpha particles, and heavy nuclei. We also partially account for pair production/annihilation
\beq
e^+ e^- \leftrightarrow \nu \bar \nu
\eeq
and nucleon-nucleon Bremsstrahlung
\beq
N +N  \leftrightarrow N+N + \nu + \bar \nu
\eeq
with $N$ any nucleon. We currently ignore reactions involving muon and tau leptons. As a result, all heavy-lepton neutrinos have the same distribution function, and we lump $\nu_\mu, \bar\nu_\mu, \nu_\tau, \bar\nu_\tau$ in a ``heavy-lepton neutrino'' species $\nu_x$.

In the calculation of these reaction rates, NuLib assumes Fermi-Dirac distributions in statistical equilibrium at the fluid density, temperature and composition for all fermions. This is very accurate for nucleons, nuclei, electrons and positrons, and reasonable for (anti)neutrinos in reactions involving only one (anti)neutrino. Indeed, neutrino blocking factors are important in hot regions, where neutrinos are close to being in equilibrium with the fluid, and negligible in colder regions, where the assumption of a thermal Fermi-Dirac distribution becomes inaccurate. That assumption guarantees that the tabulated emissivities and absorption rates will give us the correct equilibrium energy density in regions where neutrinos are in equilibrium with the fluid. 

For pair processes in low opacity regions, however, that assumption is problematic. In particular, the reaction $\nu\bar\nu \rightarrow e^+ e^-$ may be important in the low-density polar regions of merger remnants~\citep{Janka1999,Fujibayashi:2017abc}, where neutrinos are definitely not in equilibrium with the fluid. Assuming a thermal distribution of neutrinos in equilibrium with the fluid vastly underestimates the annihilation rate of neutrinos, and energy deposition in the polar regions.

The most common solution so far has been to ignore pair processes for electron-type neutrinos, and to include them approximately for heavy-lepton neutrinos, assuming a thermal distribution of heavy-lepton neutrinos in equilibrium with the fluid. This is done so that heavy-lepton neutrinos remain accurately evolved in optically thick regions. Ignoring pair processes entirely would be problematic for heavy-lepton neutrinos, as they are the only source of $\nu_x$ emission included in our simulations. Considering that all heavy-lepton neutrinos have the same distribution function in our current simulations, we do not have to worry about potential differences between the distribution functions of neutrinos and antineutrinos, and accounting for pair processes allows us to approximate their cooling effect in merger remnants. Using the same assumption for electron-type neutrinos would be more problematic, particularly in regions where the energy density of $\nu_e$ and $\bar \nu_e$ are out of equilibrium and very different from each other. One might in particular annihilate more neutrinos than antineutrinos (or vice-versa), leading to unphysical changes in the electron fraction of the fluid. 

With Monte-Carlo transport, we can do better than this. We have now implemented a more detailed computation of pair annihilation processes that should capture energy deposition in the low-density polar regions. This algorithm is described in Sec~\ref{sec:pairs}. When using this algorithm, we construct NuLib tables that include pair processes for heavy-lepton neutrinos only, as in our previous simulations. The rate of pair creation/annihilation is calculated separately in low-density regions were pair annihilation dominates over pair creation, accounting for the true distribution function of neutrinos in these regions.

The output of the NuLib library, with the options listed in this section, is a 4D table for $\eta$, $\kappa_a$, and $\kappa_s$ as a function of the fluid density $\rho$, fluid temperature $T$, fluid electron fraction $Y_e$, and neutrino energy $\nu$. The discretization in energy bins is made so that $\eta$ is the total emissivity for all neutrinos within a given energy bin, and $\eta/\kappa_a$ is the equilibrium energy density for all neutrinos within that bin. 

\subsubsection{Emission}

We emit Monte-Carlo packets using the following assumptions:
\begin{itemize}
\item Emission is isotropic in the fluid frame, and homogeneously distributed within a cell {\it in the coordinates of the simulation}.
\item During a time step, the emissivity is assumed to be constant. The time of emission of neutrinos is randomly drawn from a uniform distribution in time, and emitted neutrinos are then evolved until the end of the current time step
\item All neutrinos within an energy bin are emitted with the energy of the center of that bin in the fluid frame.
\item Emissivities at intermediate values of the fluid quantities are calculated by performing 3D linear interpolations in $\ln{\rho}, \ln{T}, Y_e$.
\end{itemize}
Spatial homogeneity of the emission is certainly an approximation, particularly for non-Cartesian grids, or in regions where the spacetime metric varies rapidly. A more accurate but costlier method would be to distribute neutrino emission equally between regions of identical {\it proper volume} in the fluid frame. The use of the central value of an energy bin for all neutrinos emitted in that bin, was suggested by~\cite{richers:15}. That choice, combined with the assumptions made when generating tables for $\eta$ and $\kappa_a$, guarantees that the equilibrium energy density of neutrinos is consistent with the desired Fermi-Dirac distribution, up to changes in the fluid-frame energy of the neutrinos as they evolve within a cell. Both assumptions are just discretization choices, and lead to errors that will converge away with increased spatial and energy resolution in simulations.

As our tables are typically inaccurate at low temperature, we also correct the emissivity for $T<0.5\,{\rm MeV}$, using 
\beq
\eta(T<0.5\,{\rm MeV}) = \eta(T=0.5\,{\rm MeV})\left(\frac{T}{0.5\,{\rm MeV}}\right)^6.
\eeq

If $\eta$ is the emissivity for a given neutrino species and energy bin in a chosen grid cell of volume $V$, and $\Delta t$ is the time step of the Monte-Carlo algorithm, then the total energy of the emitted neutrinos of that species and in that energy bin is
\beq
E_{\rm tot} = \sqrt{-g} V \Delta t \eta
\eeq
(if $\eta$ is already integrated over solid angle). If the desired energy of neutrino packets within this cell is $E_{\rm target}$ and the central value of the neutrino energies in our energy bin is $\nu$ we emit, on average, $E_{\rm tot}/E_{\rm target}$ packets, each representing $E_{\rm target}/\nu$ neutrinos. However, we can only emit an integer number of packets. In practice, fractional packets are thus treated probabilistically, i.e. if $E_{\rm tot}/E_{\rm target}=3.2$, we have a $20\%$ change of creating $4$ packets, and an $80\%$ chance of creating $3$ packets. The location of the packet is randomly drawn from a homogeneous distribution {\it in the coordinates of the simulation}, while the 4-momentum of the neutrinos is drawn from an isotropic distribution in the fluid frame. More specifically, we first construct an orthonormal tetrad in the fluid frame, and then set the 4-momentum of neutrinos in that tetrad to be
\beq
p^{\hat \mu}_{\rm fl} = \nu (1,\sin\theta \cos\phi,\sin\theta \sin\phi, \cos\theta).
\eeq
We draw $\cos\theta$ from a uniform distribution in $[-1,1]$ and $\phi$ from a uniform distribution in $[0,2\pi]$. In practice, in our code, the transformation matrix between the orthonormal tetrad in the fluid frame and the grid coordinates is only computed once per grid cell and per time step, and reused for every emission and scattering within that cell. The ability to reuse this transformation matrix is another important advantage of considering the metric and fluid variables as constant within a grid cell.

The most important choice to make here is $E_{\rm target}$, the desired energy of each packet. This energy does not need to be the same everywhere on the grid, or constant in time. Accordingly, it is the main tool available to us to choose how computational resources are distributed on our grid, i.e. where we produce more/less Monte-Carlo packets. The simplest choices would be a constant $E_{\rm target}$, a constant number of neutrinos per packet (constant $E_{\rm target}/\nu$), a constant number of packets emitted per time step in each cell (constant $E_{\rm tot}/E_{\rm target}$) or a variable $E_{\rm target}$ chosen to obtain a constant number of packets over the entire simulation. Either one has its advantages, but none is optimal for the merger simulations that we have performed so far. Instead, we consider that: 
\begin{itemize}
\item In low-density and low-temperature regions, where neutrino emission is largely negligible, it would be wasteful to emit a lot of neutrinos. We want to set a minimum energy for neutrino packets $E_{\rm floor}$, which will be used as the minimum energy of Monte-Carlo packets created in those regions.
\item In high-density, high-temperature regions, neutrinos are nearly in statistical equilibrium with the fluid. There, we want to avoid large fluctuations in the neutrino distribution function around this equilibrium value. The expectation value of the total energy of the packets within such a cell is, for a given species and energy bin, $e = \eta \sqrt{-g} V/\kappa_a$, with $\kappa_a$ the absorption opacity and $V$ the coordinate volume of the cell. Thus, if we want an average of $n_{c,\rm target}$ neutrino packets of a given species in a cell, each packet should have energy
\beq
E_{\rm eq} =  \frac{\sqrt{-g}V}{n_{c,\rm target}} \sum_{\rm bins} \frac{\eta}{\kappa_a}.
\eeq
In practice, in current simulations, we take $n_{c,\rm target}\sim 100$. In high-density regions, we have $E_{\rm eq}>E_{\rm floor}$. By using $E_{\rm target}=E_{\rm eq}$, we can avoid using all of our computational resources to evolve these equilibrium regions, while controlling the expected statistical noise for the energy of neutrinos in that region.\footnote{We note that $\eta/\kappa_a$ is finite (and typically small) in optically thin regions, so that $E_{\rm eq}$ remains well-behaved. In practice, we impose a floor of $10^{-70}$ on the value of $\kappa_a$, to avoid numerical issues.}
\item Finally, we would like to control the cost of simulations. The easiest way to do this is to choose  a target value $n_{p,\rm target}$ for the {\it total number of packets} over the entire simulation, $n_p$.
\end{itemize}
To merge these requirements, we proceed as follow:
\begin{itemize}
\item Choose an initial minimum packet energy $E_{\rm min}$ for each species (often, $E_{\rm floor}$).
\item As long as $\alpha^2 n_{p,\rm target}<n_p<n_{p,\rm target}$, use $E_{\rm target} = \max{(E_{\rm min},E_{\rm eq})}$. Here, $0<\alpha<1$ is a parameter that determines how often we change $E_{\rm min}$. We have so far used $\alpha=0.9$.
\item If $n_p<\alpha^2 n_{p,\rm target}$ for a given species, multiply $E_{\rm min}$ by $\alpha$ {\it for that species}. If $E_{\rm min}<E_{\rm floor}$, set $E_{\rm min}=E_{\rm floor}$.
\item If $n_p>n_{p,\rm target}$ for a given species, divide $E_{\rm min}$ by $\alpha$ {\it for that species}. Additionally, randomly select a fraction $(1-\alpha)$ of the existing packets and remove them from the simulation. The surviving packets now represents $(1/\alpha)$ times more neutrinos (and thus also $(1/\alpha)$ times more energy, as the $4$-momentum of individual neutrinos in the packet is kept constant).
\end{itemize}
We then have hot/dense regions with $n_{c,\rm target}$ packets per cell and per species, and low-density regions where $E_{\rm target}=E_{\rm min}$. The boundary between these two regions, and the value of $E_{\rm min}$, changes in order to limit the cost of the simulation. We note that if $n_{p,\rm target}$ is too low, hot regions that are strongly coupled to the fluid will have far fewer packets than the desired $n_{c,\rm target}$, and rapid fluctuations in the number of packets and momentum exchange between the fluid and the neutrinos may lead to increased errors and/or instabilities. {\it To test the convergence of the code, one should increase both the desired total number of packets and the desired number of packets in dense cells.}

\subsubsection{Absorption and Elastic Scattering}

Absorption and elastic scattering events are, in theory at least, very simple to implement in a Monte-Carlo algorithm. As for the emissivity, we obtain values of the absorption and scattering opacities $\kappa_{a,s}$ using 3D linear interpolation in $\ln \rho, \ln T, Y_e$. To obtain values of $\kappa_{a,s}$ at intermediate values of the energy of neutrinos, we  instead linearly interpolate $\log{\kappa_{a,s}}$ in $\nu$. We do this as $\kappa_{a,s}$ can vary by orders of magnitude between neighboring energy bins for the small number of bins used in our simulations so far ($12-16$ bins)\footnote{As neutrinos are always emitted with the central energy of a bin, we never need to interpolate $\eta$ in neutrino energy}. As for the emissivities, we correct the opacities at low temperature using
\beq
\kappa_{a,s}(T<0.5\,{\rm MeV}) = \kappa_{a,s}(T=0.5\,{\rm MeV})\left(\frac{T}{0.5\,{\rm MeV}}\right)^2.
\eeq

Before propagating a neutrino, we draw random numbers $r_a,r_s$ from an homogeneous distribution in $[\epsilon,1)$ (with $\epsilon=10^{-70}$). The time to the next absorption / scattering event is then
\beq
\Delta t_{a,s} = -\frac{\kappa_{a,s}p^t}{\nu}\ln{r_{a,s}} 
\eeq
with $\nu$ the energy of neutrinos in the fluid frame. This implies Poisson statistics for absorption and scattering events. As already discussed, we then consider the smallest time interval between $\Delta t_{a,s}$ and the desired time step. If $\Delta t_a$ is the smallest time interval, the packet is absorbed. It is then simply removed from the simulation. If $\Delta t_s$ is the smallest, the packet is scattered. We randomly draw a new 4-momentum with the same fluid-frame energy as the original packet, and a direction of propagation drawn from an isotropic distribution. We then continue the evolution from the scattering event, redrawing $\Delta t_{a,s}$. 

This simple process works well as long as $\kappa_{a,s} \Delta t \lesssim 1$, i.e. when individual grid cells are optically thin or semi-transparent. When $\kappa_{s} \Delta t \gg 1$, a single packet may undergo many scattering events during a single time step, while when $\kappa_{a} \Delta t \gg 1$, most packets are immediately reabsorbed by the fluid. Both of these cases create large computational costs, even though they also correspond to physical configurations where the evolution of the particles is fairly well understood: in high scattering regions, the neutrino energy density will evolve according to a simple diffusion equation, while in high absorption regions, the neutrinos equilibrate with the fluid on a timescale shorter than one time step and, in the case of binary mergers, on a timescale that is also much shorter than the dynamical time scale of our system. In optically thick regions, there is much to gain by using these known physical behaviors to modify the basic algorithm for absorption and scattering described in this section. We describe the choices made in the SpEC code in Secs.~\ref{sec:highKs}-\ref{sec:highKa}.

We note again that this manuscript only considers elastic scattering. Its extension to the explicit treatment of inelastic scattering is straightforward given a table for the effective opacity $\kappa_{s,in}(\nu_{in},\nu_{\rm out})$ providing the probability for a neutrino of energy $\nu_{\rm in}$ in the fluid frame to be scattered with final energy $\nu_{\rm out}$. The NuLib library used to generate $\kappa_a,\kappa_s$ for our simulations can already provide opacities for inelastic electron-neutrino scatterings, and one only need the total inelastic scattering opacity and the the post-scattering distribution of packet energies to determine the time to the next inelastic scattering events, and the energy of the packet after that event. Difficulties would however arise if the total scattering opacity 
\beq
\kappa_{s,in}(\nu_{\rm in})=\int d\nu_{\rm out} \kappa_{s,in}(\nu_{in},\nu_{\rm out})
\eeq
is high and $\kappa_{s,in} \Delta t \gtrsim 1$. In particular, the approximate methods described in Sec.~\ref{sec:highKs}-\ref{sec:highKa} would certainly need to be adapted in such regions. In the spirit of Sec.~\ref{sec:highKa}, one might consider an approximate total scattering opacity $\kappa_{s,el}+\kappa_{s,in}$ in the diffusion equation used in Sec.~\ref{sec:highKs} and, if an inelastic scattering occurred during a time step, draw the final energy of the packet from an appropriate distribution $f(\nu_{\rm out})${\it after the end of a time step}. This neglects the impact of changes in the energy of a packet during a time step on the opacity. Whether this would lead to accurate evolutions in a neutron star merger remnants is an open question, and will require further investigation. 

\subsubsection{Pair annihilation in low density regions}
\label{sec:pairs}

Neutrino-antineutrino pair annihilation is typically more difficult to take into account in simulations than the simple isotropic emission, absorption, and elastic scattering processes considered so far. A full accounting of pair processes in all regions of the simulations would require us to take into account the distribution functions of neutrinos and antineutrinos, and the energy distribution of electrons and positrons (for $e^+ e^-$ pair creations). While this is certainly possible to do using the information available in a Monte-Carlo simulation, we currently limit ourselves to a simpler numerical scheme that limits the cost of the calculation of annihilation cross-sections. As already mentioned, pair creation/annihilation of heavy lepton neutrinos in dense regions are computed assuming neutrinos in equilibrium with the fluid, while we ignore them for electron type neutrinos (as charged current reactions typically dominate the emission/absorption of electron-type neutrinos in neutron star mergers). In low-density regions, however, this would significantly underestimate the rate of neutrino-antineutrino pair annihilation, as the energy density of neutrinos is often much higher than the equilibrium density of neutrinos in equilibrium with the fluid. In those regions, we thus calculate pair annihilation rates using the evolved distribution function of neutrinos, but we neglect blocking factors in the calculations of the cross-section due to existing electrons and positrons in the fluid. We also neglect the mass of the electron, as typical neutrino energies are $\gtrsim 10\,{\rm MeV}$ in merger remnants.

The impact of pair annihilation on the evolution of the fluid is
\beq
\nabla_\mu T_{\rm fl}^{\mu\nu} = Q_{\rm pair}^\nu
\eeq
with $T_{\rm fl}^{\mu\nu}$ the stress-energy tensor of the fluid and $Q_{\rm pair}^\nu$ the rate of momentum deposition per unit volume. For the annihilation of a packet of neutrinos with momentum $p^\mu$ and stress-energy tensor $T^{\mu\nu}$ with a packet of antineutrinos of momentum $\bar p^\mu$ and stress-energy tensor $\bar T^{\mu\nu}$, $Q_{\rm pair}^\nu$ can be written as~\citep{1999ApJ...517..859S,Fujibayashi:2017abc}
\beq
Q_{\rm pair}^\nu = (p^\nu + \bar p^\nu) \frac{C_{\rm pair} c G_F^2}{3\pi} T^{\alpha\beta} \bar T_{\alpha\beta}.
\eeq
Here, $G_F^2=5.29\times 10^{-44} {\rm cm^2\,MeV^{-2}}$ is the Fermi constant, while
\beq
C_{\rm pair} = 1 \pm 4 \sin^2 \theta_W + 8 \sin^4 \theta_W,
\eeq
$\sin^2\theta_W = 0.23$, and the $+$ sign is for electron-type neutrinos while the $-$ sign is for heavy-lepton neutrinos.

The simple absorption opacity described in the previous section, on the other hand, implies
\beq
Q_{\rm pair}^\nu = \kappa_a^{\rm eq} (Ju^\nu + H^\nu) + \bar \kappa_a^{\rm eq} (\bar Ju^\nu + \bar H^\nu),
\eeq
with $\kappa_a^{\rm eq}$, $\bar \kappa_a^{\rm eq}$ the tabulated absorption opacities of neutrinos and antineutrinos, $J,\bar J$ their energy density in the fluid frame, and $H^\mu, \bar H^\mu$ their momentum density in that same frame.

For a single packet representing $N$ neutrinos of momentum $p^\mu$ and fluid-frame energy $\nu$, and smoothing out the distribution function over a cell of volume $V$,
\beq
Ju^\nu + H^\nu = N \frac{\nu}{\sqrt{-g}Vp^t} p^\mu.
\eeq
We thus define our absorption opacity for pair annihilation $\kappa_p$ using
\beq
\frac{C_{\rm pair} c G_F^2}{3\pi} T^{\alpha\beta} \bar T_{\alpha\beta} = \kappa_p N \frac{\nu}{\sqrt{-g}Vp^t}
\eeq
or, making use of our expression for the contribution of a single packet to $T^{\alpha\beta}$,
\beq
\kappa_p = \frac{C_{\rm pair} c G_F^2}{3\pi} \frac{p^\alpha p^\beta}{\nu} \bar T_{\alpha\beta}.
\eeq
In this expression, we can simply interpret $\bar T_{\alpha\beta}$ as the total stress-energy tensor of antineutrinos in the current cell. The 4-momentum $p^\alpha$, on the other hand, represents the momentum of a single neutrino in the packet for which we are calculating $\kappa_p$. Similarly, for antineutrinos,
\beq
\bar \kappa_p = \frac{C_{\rm pair} c G_F^2}{3\pi} \frac{\bar p^\alpha \bar p^\beta}{\bar \nu} T_{\alpha\beta}.
\eeq
As we lumped together all heavy-lepton neutrinos into a single species $\nu_x$, we should divide $\bar T_{\alpha\beta}$ by $4$ in that expression when considering $\nu_x$, as a given neutrino can only interact with $1/4$ of the neutrinos included in $\nu_x$ (e.g. $\nu_\mu$ only annihilates with $\bar \nu_\mu$, not $\nu_\mu$, $\nu_\tau$ or $\bar \nu_\tau$). 

We note that in deriving this expression, we have taken advantage of the relatively simplicity of the formula for the cross-section of $\nu\bar\nu$ annihilation obtained when ignoring blocking factor in the electron-positron phase space and the mass of the electron, and of the fact that all neutrinos in a given packet are assumed to have the same energy; we have not however truncated the expansion of $\kappa_p, \bar\kappa_p $ in moments of the neutrino distribution function; knowing the stress-energy tensor of antineutrinos is sufficient to calculate the effective absorption opacity of neutrinos for pair annihilation. Knowledge of the zeroth, first, and second moments of the neutrino distribution function is thus sufficient to calculate $\kappa_p, \bar\kappa_p $ in this approximation. We also note that, as long as we precompute $\bar T_{\alpha\beta}$ (which is done on the flight during the propagation of the Monte-Carlo packets, as discussed in Sec.~\ref{sec:coupling}), we can determine $\kappa_p$ for each Monte-Carlo packet without referring to other packets in our simulation.

Unfortunately, this estimate of $\kappa_p$ is only valid in low density regions. More specifically, it is valid in regions where we do not expect significant blocking factors for $\sim 10\,{\rm MeV}$ neutrinos. Accordingly, to avoid creating instabilities in our evolution and/or the introduction of large errors in high-density regions, we choose to suppress $\kappa_p$ in dense regions. We propose the simple prescription
\beq
\kappa_p \rightarrow \kappa_p e^{-\rho/\rho_{\rm crit}}
\eeq
with $\rho_{\rm crit}=5\times 10^{11}\,{\rm g/cm^3}$, as our assumptions do not appear to create significant errors in $\kappa_p$ below that density, and this should be sufficient to capture the impact of pair annihilation in the low-density polar regions of post-merger remnant. This prescription may however need to be revisited for different systems.

Practically, when it comes to pair annihilation, we take $\kappa_p$ into account by correcting the number of neutrinos $N$ represented by any given packet at the end of a time step. To do so, we assume that during a time step
\beq
\frac{dN}{dt} = - \kappa_p N \rightarrow N(t) = N_0 e^{- \kappa_p (t-t_0)}.
\eeq
After a time step $\Delta t$, we thus have $N(t_0+\Delta t)=N(t_0) e^{- \kappa_p \Delta t}$.
We also use these estimates of $ \kappa_p$ and $N(t)$ to calculate the momentum deposited by each packet into the fluid as a result of pair annihilation.

Pair annihilation is thus treated differently from other absorption processes: it never destroys Monte-Carlo packets, but only changes the number of neutrinos in each packet. The main reason to destroy packets in the 'standard' absorption process is to avoid ending up with a large number of very low-energy packets coming from regions of high optical depth. This is not an issue for pair annihilation as implemented here, as it is only active in optically thin regions. Changing the number of neutrinos in a packet then results in less shot noice in the evolution of the neutrinos, without significantly increasing the number of surviving packets.

This algorithm for neutrino-antineutrino annihilation {\it was not used in our existing merger simulation} with Monte-Carlo transport; it is proposed for the first time here, and tested in Sec.~\ref{sec:pairtest}.

\subsection{High-scattering regions: Diffusion approximation}
\label{sec:highKs}

Region of high scattering opacities ($\kappa_s \Delta t_c  \gg 1$) can become problematic in a Monte-Carlo code due to the need to evolve packets for very small time steps $\sim \kappa_s^{-1}$ between scattering events, redraw the momentum of the packet in the fluid frame after each scatter, and finally transform that momentum back into the coordinate system of the simulation. An alternative is to develop a treatment of high-$\kappa_s$ regions that makes use of the fact that the energy density of neutrinos approximately evolves according to a diffusion equation. We already presented such an algorithm in~\cite{Foucart:2017mbt}, together with comparisons of our approximate method with a full treatment of elastic scattering (i.e. evolutions in which every scattering event is treated individually). We review this method here for completeness. We also note that an alternative treatment of high-$\kappa_s$ regions in relativistic simulations that also relied on solutions of the diffusion equation was previously published in~\cite{richers:15}.

In the diffusion approximation and in flat space, the probability that a packet propagates a distance $r_d$ from its original position within a time interval $\Delta t'$ (in the fluid frame) is
\beqn
f(\tilde r) &=& \frac{4}{\sqrt{\pi}} \tilde r^2 \exp{\left(-\tilde r^2\right)},\\
\tilde r &=& \sqrt{\frac{3\kappa_s \Delta t'}{4}} r_d.
\eeqn
This solution is accurate for $\kappa_s \Delta t' \gg 1$. For an object stationary in the fluid frame, the fluid-frame time interval can be determined from the coordinate time interval $\Delta t$ using $\Delta t' = \Delta t / u^t$; this will however not be the case for packets with a non-zero average velocity in the fluid frame.

In~\cite{Foucart:2017mbt}, we argued that this formula can be made accurate for $\kappa_s \Delta t' \gtrsim 3$ if we explicitly correct it to account for the known probability that a packet experiences no scattering event during the interval $\Delta t'$,
\beq
p({\rm no\,scatter}) = \exp{(-\kappa_s \Delta t')}.
\eeq
To do this, we define the probability distribution for the value of $\tilde r$ after a time $\Delta t'$ as
\beq
p(\tilde r) = f(\tilde r) \frac{1-\exp{(-\kappa_s \Delta t')}}{\int_0^{\tilde r_{\rm max} }f(\tilde r') d\tilde r'}
\eeq
with
\beq
\tilde r_{\rm max} = \sqrt{\frac{3\kappa_s \Delta t'}{4}} \Delta t',
\eeq
an expression valid for $\tilde r<\tilde r_{\rm max}$. We also have a probability $p({\rm no\,scatter})$ that no scattering occurs, which leads to $r_d = \Delta t'$ and $\tilde r=\tilde r_{\rm max}$ (in units where $c=1$).

In practice, we tabulate the function $\tilde r(x)$ defined implicitly by the equation
\beq
\int_0^{\tilde r(x)} p(\tilde r') d\tilde r' = x,
\eeq
with $x\in [0,1-p({\rm no\,scatter})]$. We then evolve packets by drawing a number $P$ from a uniform distribution in $[0,1)$. If $P>1-p({\rm no\,scatter})$, the packet moves by $r_d = \Delta t'$. Otherwise, it moves by
\beq
r_d = \tilde r(P) \sqrt{\frac{4}{3\kappa_s \Delta t'}}.
\eeq

In our code, we always begin by propagating a packet to the first scattering event (if any), and performing the first elastic scattering. After that scattering event, and if $\kappa_s \Delta t' > 3$ for the remaining evolution time $\Delta t'$ in the fluid frame, we use the approximate diffusion method. We calculate the distance $r_d$ over which a packet moves in the fluid frame, and from there derive $f_{\rm free}=r_d/\Delta t'$. We then split the evolution of the packet into two steps: advection with the fluid for a time $\Delta t'_{\rm adv}=\Delta t' (1-f_{\rm free})$, and then free-streaming for a time $\Delta t'_{\rm free}=\Delta t' f_{\rm free}$.

To advect a packet, we first define $p^\mu = A t^\mu + B u^\mu$ with $A,B$ chosen so that $p^\mu p_\mu=0$ and $p^\mu u_\mu = -\nu$. A packet moving along a null geodesic defined by $p^\mu$ will move to the same spatial position as an observer comoving with the fluid for $\Delta t'_{\rm adv}$ after a (shorter) time $\Delta t'_{\rm com}=\Delta t'_{\rm adv} (A+Bu^t)/(Bu^t)$, at least up to changes in $p^\mu$ during the evolution of the packet along a null geodesic.\footnote{Note that $A<0$ and thus $\Delta t'_{\rm com}<\Delta t'_{\rm adv}$. Additionally, $p^\mu u_\mu = -\nu$ at the beginning of this step, but not necessarily at the end of the step; the geodesic equation determined the evolution of $p^\mu$.} We thus evolve a packet along that null geodesic for $\Delta t'_{\rm com}$, then evolve the packet along $t^\mu$ (i.e. keep it stationary in the coordinate of the simulation and keep $p^\mu$ fixed) for $(\Delta t'_{\rm adv}-\Delta t'_{\rm com})$.\footnote{If the fluid is at rest in the grid frame, i.e. $u^\mu = C t^\mu$ for some constant $C$, these equations are not well-defined but we can simply choose $\Delta t'_{\rm com}=0$. Practically, we make that choice when the grid frame speed of the fluid is below $10^{-10}c$.} This is unfortunately not a truly covariant algorithm, but it does transport packets to the correct location, and will capture changes in the energy of the neutrinos due to e.g. a gravitational redshift, as changes in the position of the neutrinos are performed by evolving packets along a null geodesic. During this advection process, we assume that the packet is stationary in the fluid frame, so that all primed (fluid frame) time intervals are related to unprimed (simulation frame) time intervals by $\Delta t' = \Delta t / u^t$.

For the free-streaming step, we first define a free-streaming 4-momentum such that $p^\mu u_\mu=-\nu$, with orientation drawn from an isotropic distribution in the fluid frame. We then propagate the packet along a null geodesic for a time $\Delta t'_{\rm free}$ in the fluid frame. Ignoring changes in $p^\mu$, $u^\mu$ and the metric, we have $\Delta t_{\rm free} = \Delta t'_{\rm free} p^t/p^{t'} = \Delta t' f_{\rm free} p^t/p^{t'}=\Delta t f_{\rm free} p^t/(u^t \nu)$. In the simulation frame, we thus evolve the packet for $\Delta t_{\rm free}$ along a null geodesic. As a result, packets propagating in different directions are evolved for a different amount of time. This is crucial if we want the packets to have, on average, zero-velocity in the fluid frame: as we drew the direction of propagation from an isotropic distribution in the fluid frame, we should also evolve packets evolving in different directions for an equal amount of time in the fluid frame, and not in the simulation frame. Practically, this means that at the end of this process, a packet may evolve by more/less than the originally requested $\Delta t$ in the simulation frame. This is not a major issue for our algorithm, however, as each packet keeps track of its own evolution time. If after such a step a packet reaches a time that is more than $0.05\Delta t_c$ earlier than the desired end time, we continue propagating it immediately; otherwise, we wait for the next call to the Monte-Carlo algorithm to do so.

Finally, we need to determine the 4-momentum of the packet at the end of the time step. If we were truly in a region with $\kappa_s \Delta t'  \gg 1$, this would be trivial; we could simply draw the 4-momentum from an isotropic distribution in the fluid frame. However, this is not the case for $\kappa_s \Delta t'  \sim 1$. In particular, any packet that, according to our algorithm, did not experience any scattering event should clearly use as final momentum the 4-momentum used during the free-streaming step of the algorithm. Any packet for which $f_{\rm free}\sim 1$ should also have a higher probability of being aligned with the 4-momentum of the free-streaming step than any other random direction. In~\cite{Foucart:2017mbt}, we derived an accurate semi-analytical model for the choice of the final 4-momentum, calibrated on simulations that treat each scattering event individually. Our method relies on the determination of an angle $\theta_2$ between the final 4-momentum $p^{\mu'}_{\rm final}$ and the 4-momentum used during the free-streaming step $p^{\mu'}_{\rm free}$ (measured in the fluid frame), as well as an angle $\phi_2$ allowing us to rotate $p^{\mu'}_{\rm final}$ around $p^{\mu'}_{\rm free}$. The angle $\phi_2$ is drawn from a uniform distribution in $[0,2\pi]$, by symmetry. Our model then uses
\beq
\cos\theta_2 = B(f_{\rm free}) - \left[1+B(f_{\rm free})\right] \exp{\left[r \ln{\frac{B(f_{\rm free})-1}{B(f_{\rm free})+1}}\right]}\nonumber
\eeq
with $B(f)$ a fitting function given in~\cite{Foucart:2017mbt} and $r$ drawn from a uniform distribution in $[0,1)$. This choice comes from the observation that the distribution of $\theta_2$ in full scattering simulations seems to mostly depend on $f_{\rm free}$. We will have $\theta_2\rightarrow 0$ when $f_{\rm free}\rightarrow 1$ if $B(1)\rightarrow 1$, and an isotropic distribution of momenta for $f_{\rm free}\rightarrow 0$ if $B(0)\rightarrow \infty$. There is otherwise no theoretical justification for this formula; it is only a semi-analytical model that matches well the results of more detailed simulations. A table for $B(f_{\rm free})$, as well as tests of this algorithm and examples of the errors that can be created if using simpler methods to calculate $p^{\mu'}_{\rm final}$ can be found in~\cite{Foucart:2017mbt}. Ultimately, the very good agreement found for the diffusion rate of neutrinos between our approximate scheme and more detailed calculations is the best test of the accuracy of our method. At the moment, we have encountered larger error in the diffusion rate of neutrinos due to the spatial discretization of the fluid evolution (and thus of the opacities) than due to any approximation made in high-$\kappa_s$ regions. Increasing the minimal value of $\kappa_s \Delta t'$ above which we use this diffusion approximation has a minimal impact on the result of the evolution in our existing tests.

\subsection{High absorption regions: Implicit Monte-Carlo}
\label{sec:highKa}

The treatment of regions with high absorption opacities ($\kappa_a \Delta t \gg 1$) is probably the most important challenge faced when using Monte-Carlo methods to evolve neutrinos in neutron star merger simulations. In our first uses of Monte-Carlo transport, we did not need to evolve these regions directly; the Monte-Carlo code was used to close the two-moment equations~\citep{Foucart:2017mbt} or without direct coupling to the fluid equations~\citep{Foucart:2018gis}. For the purpose of Monte-Carlo transport, we then simply assumed statistical equilibrium with the fluid in dense regions. 

Coupling the moment formalism with Monte-Carlo methods does however have important disadvantages, most importantly that the two evolution methods may produce diverging solutions resulting in unphysical artifacts~\citep{Foucart:2017mbt}. When attempting to use Monte-Carlo methods coupled to a two-moment scheme in merger simulations, we also found problematic violations of the conservation of energy and lepton number in the intermediate regions where we transition from moments-only evolution (dense regions) to Monte-Carlo closures (low-density regions). It may very well be possible to resolve these issues with a more careful coupling of the two methods, but a Monte-Carlo-only evolution of the neutrinos allows us to automatically avoid these issues. 

Monte-Carlo methods however have their own drawbacks in these optically thick regions: the average packet will only survive for a time $\sim \kappa_a^{-1}$, and thus if $\kappa_a \Delta t \gg 1$, most packets created during a neutrino time step are immediately reabsorbed. This is quite wasteful, as at the same time we can reasonably expect these packets to sample a relatively simple equilibrium distribution function; computational resources would be more usefully spent on packets in the harder-to-model semi-transparent regions. When $\kappa_a \Delta t \gtrsim 1$, simple explicit time stepping methods can also run into stability issues.

To get a more efficient algorithm, we note that the absorption of a neutrino followed by the emission of a neutrino of the same energy at the same point is practically identical to an elastic scattering event. If neutrinos in a given energy bin are exactly in equilibrium with the fluid, then the transformation
\beq
\eta' = \alpha \eta;\,\, \kappa_a' = \alpha \kappa_a;\,\, \kappa_s' = \kappa_s + (1-\alpha) \kappa_a
\eeq
leaves the neutrino distribution unmodified. This is no longer exactly the case after spatial discretization of the problem or when neutrinos are out of equilibrium with the fluid, yet
 rigorous Implicit Monte-Carlo (IMC) methods can be developed based on this idea. In particular, if $\alpha$ takes the same value for all energy bins and remains within a specific range,
and if the implicit scattering opacity $\kappa_s = (1-\alpha)\kappa_a$ represents {\it inelastic} scatterings creating packets with an isotropic distribution of momenta and a thermal distribution of energies,
then this transformation is equivalent to a well-chosen time-discretization of the problem~\citep{1971JCoPh...8..313F}.

In this manuscript, we rely on a more aggressive approximation. We assume that the above transformation will remain reasonable as long as (a) it is performed well inside of the neutrinosphere; (b) $1/\kappa_a'$ is small compared to the dynamical time scale of our system; and (c) $1/\sqrt{\kappa_a'(\kappa_a'+\kappa_s')}$ is small compared to the length scale over which the fluid and neutrino distribution function vary. The first condition implies a quasi-equilibrium distribution function of neutrinos, while the second and third mean that the distribution function does not vary significantly between two absorption events, after transformation to the new emissivities and opacities. Practically, we choose $\alpha$ so that $\kappa_a' \Delta t_c \leq \xi$ (we typically choose $\xi=1$, though the exact value can be specified at run time). The first two conditions can then be restated as a requirement that cells with optical depth $\tau=\kappa_a' \Delta t_c$ are inside the neutrinosphere (true in our simulations so far), and that the light crossing time of a grid cell is small compared to the dynamical time scale of the system (always true in binaries). Our transformation then leaves the equilibrium energy density of neutrinos $\eta/\kappa_a$ and the diffusion timescale $1/(\kappa_a +\kappa_s)$ unmodified, while increasing the equilibration and thermalization time scales so that they are at least of the order of the light-crossing time of a grid cell.

To avoid instability in the coupled evolution of the fluid and neutrino, we also borrow from IMC methods and define
\beq
\beta = \max{\left(\left|\frac{d u_\nu}{d u_{\rm fl}}\right|, m_p c^2 \left|\frac{d n_\nu}{d (\rho Y_e)_{\rm fl}}\right|\right)}
\eeq
with $u_{\nu,\rm fl}$ the neutrino and fluid equilibrium energy density, $n_\nu$ the neutrino electron number density (number density of $\nu_e$ minus number density of $\bar \nu_e$), $\rho$ the fluid baryon density, $Y_e$ the fluid electron fraction, and $m_p$ the proton mass.\footnote{$u_\nu$ and $n_\nu$ can be extracted from the tabulated values of $\eta$ and $\kappa_a$.} We take the first derivative at constant $\rho,Y_e$ and the second at constant $\rho,T$, with $T$ the fluid temperature. A large $\beta$ indicates that a small change in the temperature or electron fraction of the fluid due to neutrino emission / absorption leads to large changes in the equilibrium distribution function of neutrinos. We can get an idea of the role of $\beta$ by considering the coupled evolution of the neutrino and fluid energy densities for a homogeneous medium and in the fluid frame,
ignoring changes in the composition of the fluid:
\beqn
\frac{dJ}{dt} &=& -\kappa'_a (J-u_{\nu})\\
\frac{du_{\rm fl}}{dt} &=& \kappa'_a (J-u_{\nu}).
\eeqn
Here $J$ is the fluid-frame energy density, and $u_\nu=\eta/\kappa_a$. Defining $\delta J=J-u_{\nu}$ and combining these equations, we get
\beq
\frac{d\delta J}{dt} + \tilde \beta \kappa'_a \delta J = -\kappa'_a \delta J
\eeq
with $\tilde \beta=du_\nu/du_{\rm fl}$. For a standard forward-Euler explicit time stepping scheme, the stability condition is then
\beq
\kappa'_a \Delta t \leq \frac{1}{1+\tilde \beta},
\eeq
or
\beq
\alpha < \frac{1}{1+\tilde \beta} \frac{1}{\kappa_a \Delta t}.
\eeq
The same argument holds for the evolution of the electron lepton number ignoring the evolution of the fluid energy density, except that we need to take $\tilde \beta = m_p dn_\nu/d(\rho Y_e)$. 
Practically, we impose
\beq
\alpha < \frac{1}{1+\beta} \frac{1}{2C\kappa_a \Delta t_c}
\eeq
with $C$ the minimum Courant factor used by the Monte-Carlo algorithm. For our typical values of $C=0.5$ and $\xi=1$, and given that $\beta>0$, this condition is always more restrictive than the condition $\kappa_a'\Delta t_c < \xi$; but usually not by much as in most regions of our simulations $\beta$ is small. When imposing smaller values of $\xi$ or using a smaller Courant factor, the first condition may become more restrictive.

There are of course caveats to this derivation. The first is that we do not truly use forward-Euler time stepping. This would assume that the neutrino energy density $J$ in the right-hand-side of these equations is set to its value at the beginning of a time step, while in a Monte-Carlo algorithm it is actually an up-to-date value of $J$ set by the actual number of Monte-Carlo packets on the grid. A more careful study of our time stepping algorithm shows that the condition imposed in our code is more restrictive than strictly required. The second is that using derivatives at constant $Y_e$ or $T$ to calculate $\beta$ is an approximation, and does not necessarily return the maximum potential value of $\beta$. Derivatives along the actual trajectory of the fluid in the $(\rho,T,Y_e)$ parameter space would be preferable, but would require an implicit solve of both the fluid equations and the transport equations. Finally, we ignored the impact of spatial inhomogeneities on the stability of our system. 

As more computational resources become available, we should be able to jointly decrease the grid spacing and time step. The maximum value of $\kappa_a'$ then increases, limiting the impact of our approximation. Increasing $\kappa_a'$ without decreasing the grid spacing and time step, on the other hand, would require a more careful implicit treatment of the coupling between the fluid and the neutrinos. This would certainly be more expensive, but necessary to get order-of-magnitude increases in $\kappa_a'$.

Tests of this approximate treatment of high-$\kappa_a$ regions are presented below, and are ultimately the main indication that this method provides reasonable results. We can however get a rough idea of the error that they create by considering the two-moment equation for the energy density $J$ and momentum density $H^\mu$ of neutrinos in optically thick regions. With planar symmetry and in a coordinate system where the fluid is at rest, we get (in units with $c=1$)
\beqn
\partial_t J + \partial_x H &=& \eta -\kappa_a J\\
\partial_t H +\frac{1}{3} \partial_x J &=& -(\kappa_a+\kappa_s) H,
\eeqn
where we used the optically thick closure $P_{ij}=\delta_{ij} J/3$ for the pressure tensor $P_{ij}$. This is equivalent to
\beqn
\partial_t J + \partial_x H &=& \eta' -\kappa_a' J + (1-\alpha) (\eta-\kappa_a J)\\
\partial_t H +\frac{1}{3} \partial_x J &=& -(\kappa_a'+\kappa_s') H.
\eeqn
If $J=J_{\rm eq} + \delta J$ with $J_{\rm eq}=\eta/\kappa_a$, then the transformation $\eta \rightarrow \eta' + \delta \eta'$ with $\delta \eta' \sim -(1-\alpha) \kappa_a \delta J= -(1-\alpha) \eta (\delta J/J_{\rm eq})$ would leave the solution unmodified. This is not what we are doing, however. In our approximation, we set $\delta \eta'=0$. The resulting relative error in the emissivity is
\beq
\left|\frac{\delta \eta'}{\eta'}\right| \sim \frac{1-\alpha}{\alpha} \frac{\delta J}{J_{\rm eq}} < \frac{\kappa_a \Delta t_c}{\xi} \frac{\delta J}{J_{\rm eq}}
\eeq
in regions where $\alpha \kappa_a \Delta t_c = \xi$.
Integrating the original energy equation in steady state over a single cell, we also get
\beq
\Delta H = -\kappa_a \Delta t_c \delta J
\eeq
with $\Delta H$ the contribution of that cell to the neutrino flux. Thus
\beq
\left|\frac{\delta \eta'}{\eta'}\right| < \frac{\Delta H}{\xi J_{\rm eq}}.
\label{eq:error}
\eeq
If only applied in optically thick regions (where $\Delta H \ll J_{\rm eq}$) and for our choice of $\xi \sim 1$, this is clearly a small correction. As the approximate equations used in this derivation are linear in $J, H$, Eq.~\ref{eq:error} is also a reasonable estimate of the error in the energy density, flux, and luminosity of neutrinos outside of the neutrinosphere. 

While this is certainly not a rigorous error estimate for more generic systems, we expect that as long as the evolution of a system is slow compared to the evolution time step, our error estimate will remain order-of-magnitude accurate. We note that the fractional correction to $\eta'$ becomes significant if this approximate algorithm is used at or outside of the neutrinosphere, or if $\xi \ll 1$. More generally, the error associated with this estimate will be larger in regions where we rapidly transition from free streaming neutrinos to optically thick cells than in regions where that transition occurs over many grid cells. How well this approximation would work in evolutions considering more advanced reactions that strongly couple the distribution function of neutrinos of different energies (e.g. in the presence of significant inelastic scattering) remains an open question.

\subsection{Neutrino-matter coupling}
\label{sec:coupling}

\subsubsection{Source terms for fluid evolution}

There have been at least two main methods suggested so far to handle the coupling of radiation to matter in Monte-Carlo simulations. In the first method, we explicitly keep track of all momentum and lepton number exchanges between the fluid and the neutrinos. In our code, this include all emission, absorption, scattering, and pair annihilation events. In the second method, we calculate the expectation value of momentum and lepton number exchanges during these events, given the available neutrino packets. In that case, we do not consider whether e.g. a packet was absorbed by the code. We instead estimate the likelihood of that absorption occurring during a time step, and derive from there the expectation value for momentum and lepton number transfers due to absorptions. The first method, implemented e.g. in~\cite{Ryan2015}, has the advantage to explicitly conserve energy, momentum, and lepton number. The second only does so on average, but reduces shot noise in the coupling terms due, for example, to the unlikely absorption of a packet in a low-density region of the fluid. In our current simulations, we use a relatively low number of packets and we are thus likely to be hurt by shot noise in the coupling terms. Accordingly, we use the second method. We do however keep track of momentum and lepton number transfers using the first method, and verify that the two methods agree if the coupling terms are integrated over a sufficiently long period of time.

In the SpEC code, we evolve the fluid variables
\beqn
\rho_* &=& \rho \sqrt{\gamma}\alpha  u^t\\
\tilde \tau &=& \sqrt{\gamma} T^{\mu\nu}_{\rm fl} n_{\mu} n_{\nu} - \rho_*\\
\tilde S_i &=& -\sqrt{\gamma}T^{\mu\nu}_{\rm fl}u_\mu \gamma_{\nu i} 
\eeqn
and $\rho_* Y_e$, with $n_\mu = (-\alpha,0,0,0)$ the unit one form to constant-$t$ slices and $\gamma$ the determinant of the spatial metric. The source terms appearing in the evolution of these variables due to neutrino-matter interactions are
\beqn
\partial_t \tilde \tau &=& ...+ \alpha \sqrt{\gamma} S^\alpha n_\alpha\\
\partial_t \tilde S_i &=& ... -\alpha \sqrt{\gamma} S^\alpha \gamma_{\alpha i}\\
\partial_t (\rho_* Y_e) &=& ... - \sum s_i m_p\alpha \sqrt{\gamma} \frac{\eta-\kappa_a J}{\nu}
\eeqn
where in the last term the sum is over all species and energy bins, and $s_i=1$ for $\nu_e$, $-1$ for $\bar \nu_e$, and $0$ otherwise. The source term is, for the interactions considered here,
\beq
S^\alpha = \sum \eta' u^\alpha - \sum \left( \kappa_a' J u^\alpha + (\kappa_a' + \kappa_s') H^\alpha\right ).
\eeq
The first sum is over all species and energy bins, and the second over all packets. Calculating the changes in the evolved fluid variables over one time step $\Delta t$ thus requires calculations of
\beq
\int \kappa'_a J dt ;\,\, \int (\kappa_a' +\kappa_s') H^\alpha dt;\,\, \int \kappa_a' J dt/\nu,
\eeq
with the integral taken over the current time step. Whenever a packet is evolved along a null geodesic, we calculate its contribution to these source terms following the method of Sec.~\ref{sec:mom}.  

We can verify that this provides us with the correct expectation value for energy transfers. The probability distribution for the time $t$ that will pass before a packet is absorbed is, in the fluid rest frame,
\beq
p_a(t) = \kappa_a e^{-\kappa_a t}
\eeq
and thus the expectation value for energy deposition by a packet with fluid frame energy $J_0$ in a cell with absorption opacity $\kappa_a$ over a time $\Delta t$ is
\beq
\langle \Delta J \rangle = \int_0^{\Delta t} dt p_a(t) J_0 = J_0 \left( 1-e^{-\kappa_a \Delta t}\right).
\eeq
Our code instead adds to the fluid an energy $\kappa_a J_0 \min{\left(\Delta t_a,\Delta t\right)}$, with $\Delta t_a$ the time to the actual absorption of the packet. The expectation value for the energy deposited is then
\beqn
\langle \Delta J \rangle &=&\int_0^{\Delta t}  dt \kappa_a e^{-\kappa_a t} \kappa_a J_0 t + e^{-\kappa_a \Delta t} \kappa_a J_0 \Delta t\\
&=& J_0 \left( 1-e^{-\kappa_a \Delta t}\right).
\eeqn
The two methods thus have the same expectation value for energy deposition, as desired. The same derivation can be performed for linear momentum and lepton number exchanges.

For packets that are advected with the fluid, we perform these calculations assuming that $H^\alpha=0$ and $p^\mu = \nu u^\mu$, which is correct in the average. Finally, for pair annihilation we add to the absorption term the contribution of the correction $\Delta \kappa_a$ computed in Sec.~\ref{sec:pairs}. For example, at the end of a time step $\Delta t$, a packet within a cell of volume $V$ that initially represents $N_0$ neutrinos subjected to a correction $\Delta \kappa_a$ contributes an additional term to the integral of $\kappa_a J$:
\beqn
\left(\int dt \kappa_a J\right)_{\rm pairs} &=& \int \left(N_0 e^{-\Delta \kappa_a t}\right) \frac{\Delta \kappa_a  \nu^2_k}{\sqrt{-g}V p^t_k}dt \\
&=& \left(1-e^{-\Delta \kappa_a\Delta t}\right) \frac{N_0 \nu^2_k}{\sqrt{-g}V p^t_k}.
\eeqn
We can use these expressions to consistently calculate the expectation value for energy and momentum transfer between the neutrinos and the fluid.

\subsubsection{Coupling shot noise}
\label{sec:shotnoise}

Let us now consider what sets the level of shot noise in our estimates of neutrino-matter interactions, particularly in optically thick regions where $\kappa_a' \Delta t_c = \xi$ (the maximum value allowed for the effective absorption coefficient $\kappa_a'$ within a cell). During a time step $\Delta t \lesssim \Delta t_c$, shot noise in the emissivity $\eta'$ is inexistent, as we use the tabulated values of $\eta'$ rather than the energy of the emitted packets to calculate source terms proportional to $\eta'$. Shot noise in scattering terms will typically be small as soon as $\kappa_s' \Delta t_c \gg 1$, as packets are then mostly advected with the fluid and source terms proportional to $\kappa_s'$ are explicitly set to their true expectation value ($0$) during packet advection. Source terms proportional to $\kappa_a'$, on the other hand, will be dominated by the term proportional to $\kappa_a' J$. During a time step, the shot noise in $\kappa_a' J \Delta t$ can be estimated as
\beq
\kappa_a' \Delta t \delta J_{\rm MC} = \xi \frac{J}{\sqrt{n_{c,\rm target}}} \frac{\Delta t}{\Delta t_c}
\eeq
with $n_{c,\rm target}$ the expected number of packets in the cell. We thus see that, when using a constant Courant factor $\Delta t/\Delta t_c$, multiplying $\xi$ by a factor $\alpha$ requires us to multiply $n_{c,\rm target}$ by a factor $\alpha^2$ in order to keep the same shot noise in neutrino-matter interactions in the highest $\kappa_a'$ cells (i.e. where instabilities, or even a slow random-walk motion away from the true solution, are the most likely to occur). This shows that increasing the maximum value of $\kappa_a'$ (i.e. increasing $\xi$) comes at a steep computational cost if we want to avoid shot noise in optically thick regions! 

We note that the issue of shot noise in optically thick regions is separate from the instabilities in neutrino-matter coupling that motivated some limits placed on $\kappa_a'$ in Sec.~\ref{sec:highKa}. The limits placed in that section aimed to keep the system of equations stable in the continuum regime, while shot noise is an issue due to the finite number of packets used to represent the neutrino distribution function. We can now see an important trade-off made when using our Monte-Carlo algorithm: increasing $\kappa_a'$ decreases the error introduced by correcting the equation of radiation transport in optically thick regions (replacing absorptions by elastic scattering); however, increasing $\kappa_a'$ introduces more shot noise in any given time step of the evolution, which can only be decreased by either decreasing the time step or increasing the number of packets emitted within optically thick cells. Or, stated otherwise, increasing the maximum value of $\kappa_a'$ brings us closer to using the true equations in the continuum regime but, at constant computational resources, it quickly increases shot noise in the simulation. With our standard choice of $\xi\sim 1$, $n_{c,\rm target} \sim 100$, and $\Delta t \sim 0.5 \Delta t_c$ we get shot noise of $\sim 0.05J$ for the energy absorption per time step and per unit volume. 

\subsection{Parallelization}

One of the main issue that we face when using Monte-Carlo methods for radiation transport is maintaining proper scaling of the algorithm. Indeed, Monte-Carlo packets are far from homogeneously distributed. This is by design: Monte-Carlo algorithms allow us to use computational resources where they are most needed, by placing most Monte-Carlo packets in the regions where neutrinos are important to the evolution of the fluid. Nevertheless, this means that we cannot maintain good scaling if we evolve Monte-Carlo packets on the processor responsible for the evolution of the grid cell that contains them.

In SpEC, we can `loan' {\it all packets within a given fluid cell} to a different processor, to improve load balancing. We proceed as follow:
\begin{itemize}
\item During a time step, we keep track of the number of packets evolved within each cell of the fluid grid, whether that packet is absorbed, scattered, or free-streaming. The number of packets within a cell is considered to be the `cost' of that cell.
\item At the end of a time step, we check whether load-balancing is required. We attempt to keep the cost of Monte-Carlo evolution on each processor (as defined above) below $125\%$ of the average cost on a processor. If a processor has a cost above that threshold, we try to improve load-balancing by removing packets from the processor with the highest estimated cost. This is done by, in order of priority, (i) Sending back to the processor responsible for the evolution of the fluid any packet that was previously loaned to the costlier processor, to limit communication costs; (ii) Loaning packets on fluid cells evolved by the costlier processor to the processor with the lowest estimated load, starting with the highest cost cells. To avoid complicating the communication pattern, we forbid loans of cells located within one cell width of the boundary of the domain, or within one cell width of a fluid cell evolved by another processor.
\end{itemize}
Loaned cells require the following communications between processors, all performed using asynchronous MPI communication between the processor owning the fluid data and the processor evolving the Monte-Carlo packets:
\begin{itemize}
\item Communication of the fluid and metric variables from the processor evolving the fluid to the processor evolving Monte-Carlo packets, before evolution of the packets. We need information about any loaned cell, as well as any immediate neighbor of those cells (as the evolution of Monte-Carlo packets require one layer of `ghost' cells).
\item Communication of packets that moved from one fluid cell to another to the processor now responsible for their evolution. As there is no global communication of information between all processors in our algorithm, this is a three steps process. First, packets that started the current time step in a fluid cell that is owned by processor $A$ and loaned to processor $B$, yet ended their time step in a cell that is not loaned to $B$ are sent from $B$ back to $A$. After this step, all packets that should be moved from one processor to another are owned by the processor that owns the fluid cell where they {\it started} the current time step. Second, packets that are currently on processor $A$ but moved to a fluid cell owned by processor $B$ are communicated from $A$ to $B$. After this step, all packets  that should be moved from one processor to another are on the processor that owns the fluid cell in which they {\it ended} the current time step. Finally, all packets within a fluid cell owned by processor $A$ but loaned to processor $B$ that are current owned by $A$ are sent to $B$. All packets are then on the processor responsible for their Monte-Carlo evolution. 
\item Communication of all integrated moments of the neutrino distribution function needed for coupling between neutrinos and the fluid and/or pair annihilation calculations from the processor evolving the Monte-Carlo packets (including its ghost cells) to the processor evolving the fluid
\item Communication of all integrated moments of the neutrino distribution function between neighboring regions of the fluid grid evolved on different processors. This is required because packets may deposit energy-momentum in either their current cell or immediate neighbors of that cell.
\item If using pair annihilation, communicate all moments needed for the calculation of $\Delta \kappa_a$ from the processor evolving the fluid to the processor evolving MC packets, before the calculation of $\Delta \kappa_a$.
\item If using pair annihilation, send back to the processor evolving the fluid information about energy-momentum deposition from pair annihilation, after calculation of $\Delta \kappa_a$ for each packet.
\end{itemize}

We see that proper parallelization of the Monte-Carlo codes requires a significant amount of bookkeeping and communication infrastructure, with a very different communication pattern from other methods used in general relativistic merger simulations. While in theory simpler than many other parts of the algorithm described in this manuscript, the practical implementation of this communication infrastructure is one of the main difficulties in the creation of a Monte-Carlo neutrino transport code for neutron star merger simulations. Further improvements to this algorithm are possible: for the relatively low resolution merger simulations with Monte-Carlo transport presented in~\cite{Foucart:2020qjb}, about $50\%$ of the computational time was spent in the Monte-Carlo algorithm, with $\sim 2:1$ ratio between the highest computational load and the average computational load. This is much better than if packets are evolved on the processor responsible for the fluid evolution, but certainly leaves room for future improvements.

\section{Code tests}
\label{sec:tests}

We now move to tests of our Monte-Carlo implementation. We focus on a series of simple tests for individual aspects of our algorithm, as well as a few more challenging shock tube tests and a more complex test for the evolution of neutrinos partially coupled to the fluid that uses initial conditions based on the result of a core-collapse simulation. We note that additional tests of many aspects of this algorithm can be found in~\cite{Foucart:2017mbt}, including detailed tests of the behavior of our code in regions of high scattering opacity, and of neutrino advection by the fluid. As the treatment of scattering in our simulations has not changed since that publication, we do not repeat these tests here. A short neutron star merger simulation (ending $5\,{\rm ms}$ after merger) using Monte-Carlo transport was also presented in~\cite{Foucart:2020qjb}.

\subsection{Emitting Sphere: Relativistic Beaming}

\begin{figure}
\centering
\includegraphics[width=0.5\columnwidth]{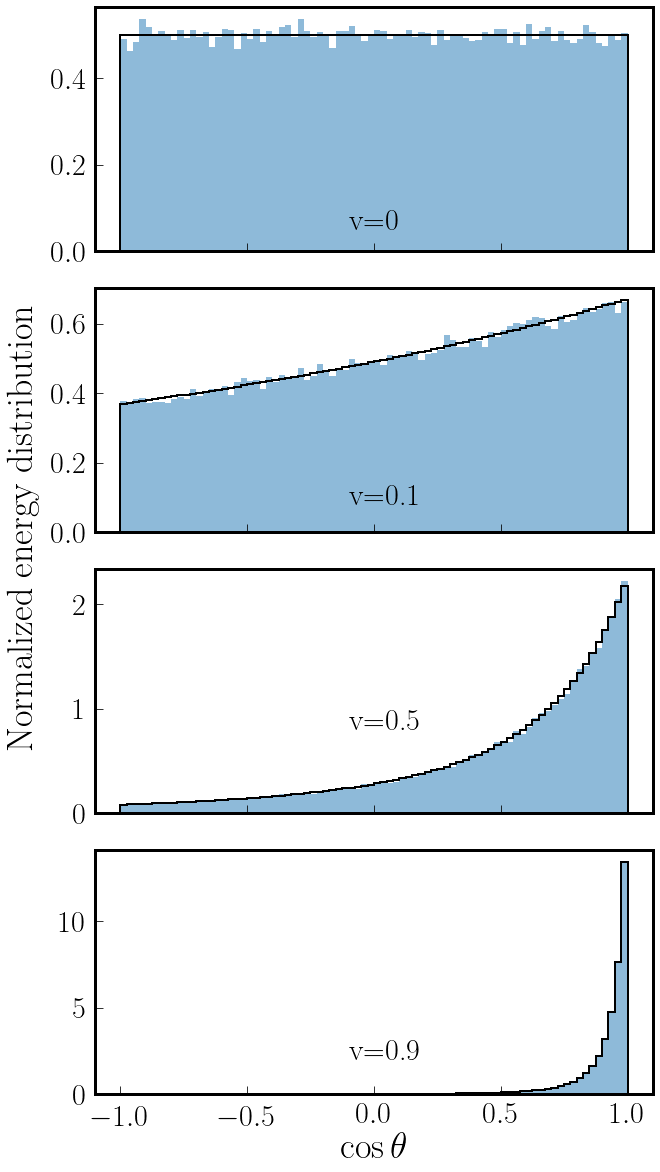}
  \caption{Normalized energy distribution of neutrinos for the emitting sphere test. We show numerical results (blue bins) and analytical expectations (black lines) for emission frames boosted by $v/c=0,0.1,0.5,0.9$ (from top to bottom). Fig.~\ref{fig:EmittingSphereError} shows that errors in this plot are consistent with the expected statistical noise from Monte-Carlo methods.}
\label{fig:EmittingSphereAngle}
\end{figure}

We begin with a test that is quite simple for a Monte-Carlo algorithm, yet very inaccurate when using the coupled two-moments/Monte-Carlo methods that we proposed in~\cite{Foucart:2017mbt}. We consider a ball of unit radius with $\kappa_a=\kappa_s=0$ and a constant emissivity $\eta$ inside the ball. We emit packets isotropically {\it in a boosted reference frame} with velocity $v/c=0,0.1,0.5,0.9$ along the x-axis. If $\theta$ is the angle between the direction of propagation of a packet and the x-axis in the simulation frame, $\theta'$ the same angle within the boosted emission frame, and all neutrinos have energy $\nu'$ in the emission frame, then
\beqn
\cos\theta' &=& \Gamma \frac{\cos\theta-v}{\sqrt{\sin^2\theta+\Gamma^2(\cos\theta-v)^2}}\\
\nu' &=& \nu \Gamma (1-v\cos\theta)\\
\Gamma &=& \frac{1}{\sqrt{1-v^2}}
\eeqn
with $\nu$ the energy of neutrinos in the simulation frame.
The distribution function of neutrinos as a function of $\cos\theta$ is
\beq
f(\cos\theta) = f(\cos\theta') \frac{d\cos\theta'}{d\cos\theta} = \frac{1}{2} \frac{d\cos\theta'}{d\cos\theta} 
\eeq
as $f(\cos\theta')=1/2$ for isotropic emission. From there, we can calculate any projection of the neutrino distribution function. In Fig.~\ref{fig:EmittingSphereAngle},
we compare numerical results and theoretical predictions for the normalized distribution of energy as a function of $\theta$. Theoretically, we expect
\beq
g(\cos\theta) \propto \nu f(\cos\theta) =  \frac{1}{2} \left(\frac{d\cos\theta'}{d\cos\theta}\right) \frac{\nu'}{\Gamma(1-v\cos\theta)}.
\eeq
Fig.~\ref{fig:EmittingSphereAngle} shows $g(\cos\theta)$, with the results binned in $80$ equal intervals in $\cos\theta$.  
Each simulation includes $\sim 10^5$ packets. We see that the numerical results closely match the theoretical distribution function.

\begin{figure}
\centering
\includegraphics[width=0.5\columnwidth]{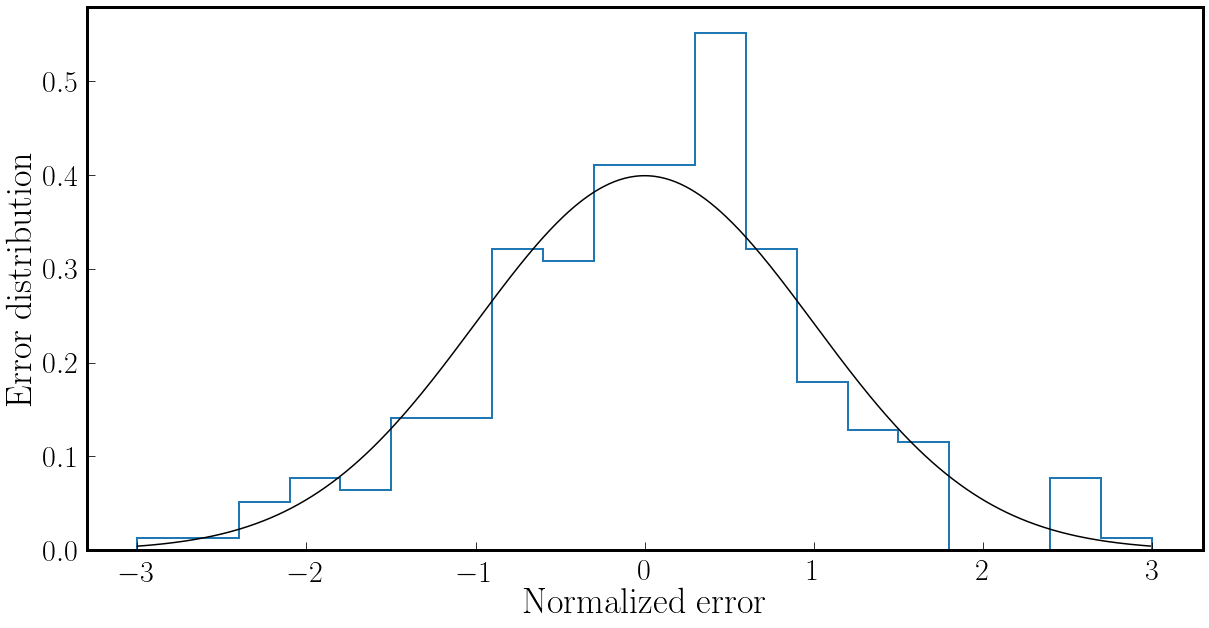}
\caption{Distribution of statistical errors for the results presented in Fig.~\ref{fig:EmittingSphereAngle}, normalized by the expected standard deviation for events following Poisson statistics (blue), and expected distribution of these errors (black, normalized Gaussian of unit width). We see that the errors are consistent with theoretical expectations.}
\label{fig:EmittingSphereError}
\end{figure}

We can go further and actually compare the observed errors with the expected sampling noise in our simulations. If a bin is filled with $N_b$ packets, the relative statistical error should be $\sim N_b^{-1/2}$ (at 1-$\sigma$). In Fig.~\ref{fig:EmittingSphereError}, we show
the distribution of statistical errors for the 320 bins available to us (80 bins $\times$ 4 simulations), normalized to their expected 1-$\sigma$ errors. We see that the resulting distribution of errors is reasonably close to the expected Gaussian.


\subsection{Propagation in the Schwarzschild metric}

For our second series of tests, we consider the evolution of neutrino packets in the Schwarzschild metric, for a black hole of mass $M_{\rm BH}$. We begin with a simple test that provides us with an estimate of the accuracy of the evolution of the 4-momentum of neutrinos: a single cell of the fluid grid has non-zero emissivity, while $\kappa_a=\kappa_s=0$ everywhere. We then verify that $p_t$ is conserved by comparing its value at the time of emission to its value for packets leaving the computational domain. Analytically, $p_t$ is a conserved quantity, but as we do not take advantage of that symmetry in our code and instead simply evolve $p_{x,y,z}$, the change in $p_t$ is a reasonable estimate of the error in $p^\mu$.

Our finite difference grid covers the region $[2.5,12.5]\times [-2,10] \times [-2.55,2.45]$ with $100\times 120\times 50$ cells. We use standard Schwarzschild coordinates with $G=M_{\rm BH}=c=1$. For $M_{\rm BH} \sim 3M_\odot$, this corresponds to a grid spacing $\Delta x \sim 450\,{\rm m}$, which is coarser that the resolution typically used in merger simulations close to a black hole or neutron star ($\sim 100\,{\rm m}-200\,{\rm m}$). We emit neutrinos from a cell centered on the point $(5.05,5.05,0)$. We then repeat this test at higher resolution (dividing $\Delta x$ by $2$). 

\begin{figure}
\centering
\includegraphics[width=0.5\columnwidth]{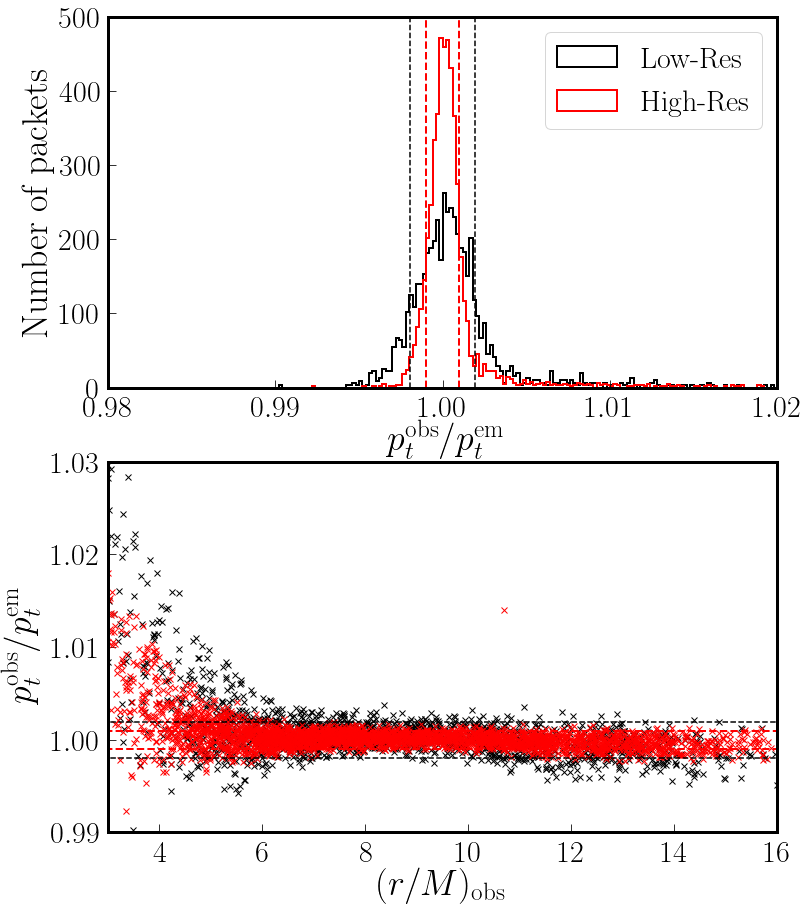}
\caption{Conservation of energy in our Monte-Carlo code, tested by propagating packets emitted with a fixed energy from a single cell of our computational domain (see text). {\it Top}: Histogram of the ratio of the conserved quantity $p_t$ for packets observed as they leave the grid to the expected value of $p_t$ for packets emitted at the center of the emitting cell. We show results for two grid resolutions, with the grid spacing changing by a factor of $2$ between resolutions. The dashed lines show the expected spread in the value of $p_t$ due to the finite size of the emitting cell. {\it Bottom}: Same ratio now shown for all observed packets, as a function of the radius at which they leave the grid. We see that outside $r\sim 6M$, the intrinsic spread in $p_t$ expected from spatial discretization errors is larger than the error in $p_t$ at the time of observation. The error in $p_t$ becomes slightly larger close to the black hole.}
\label{fig:GravitationalRedshift}
\end{figure}

The resulting errors in $p_t$ are plotted on Fig.~\ref{fig:GravitationalRedshift}. We see that for most packets, the relative change in $p_t$ is well below $1\%$. Considering that we are currently using tables for neutrino-matter interactions with $\sim 10-20$ energy bins only, that level of error in $p_t$ is unlikely to be significant. The only packets showing larger errors are the ones escaping the computational domain at low radii. We also see a clear decrease in the error when going to higher resolution, although convergence is slow. This is expected given the low-order numerical methods used in this work. 

We also see that the error in $p_t$ is comparable to the discretization error due to the use of a constant value of the metric within a cell. The true value of $p_t$ within the cell should, for constant neutrino energy in the emission frame, have a finite width, which we can calculate analytically as we know the metric and grid size. We see on Fig.~\ref{fig:GravitationalRedshift} that this width is of the same order as the final error in $p_t$. We can thus only say that the propagation error is at most comparable to the discretization error, and may very well be smaller. As a result of this test, we see that our algorithm properly captures gravitational redshifts, and we get a first indication that the propagation of packets along null geodesics is performed accurately.

\begin{figure}
\centering
\includegraphics[width=0.5\columnwidth]{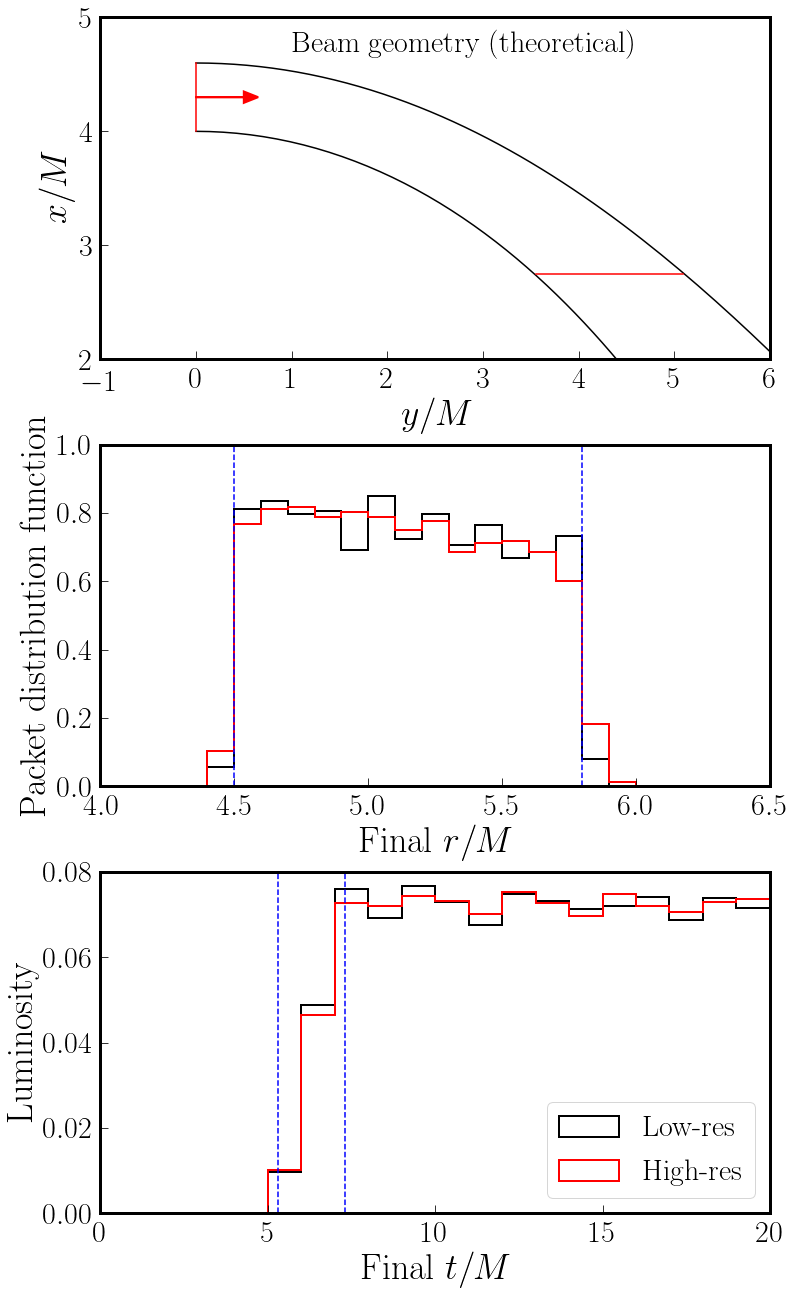}
\caption{Test of the propagation of a radiation beam in the Schwarzschild metric. {\it Top}: Geometry of the problem. The beam originates at $y=0$, and propagates along null geodesics to $x=2.75$. The red lines show where the beam is emitted and observed. {\it Middle}: Distance to the center of the black hole for Monte-Carlo packets exiting the computational grid at low (black) and high (red) resolution (10 times more packets). Dashed blue lines show theoretical expectations for the width of the beam. {\it Bottom}: Time at which Monte-Carlo packets escape the computational domain. Dashed blue lines show the range of theoretical expectations (the inner part of the beam exits the grid faster than the outer part). At late times, we expect constant neutrino luminosity. On the two lower panels, we note the decrease in sampling noise at higher resolution.}
\label{fig:PropagationSchwarzschild}
\end{figure}

\begin{figure}
\centering
\includegraphics[width=0.5\columnwidth]{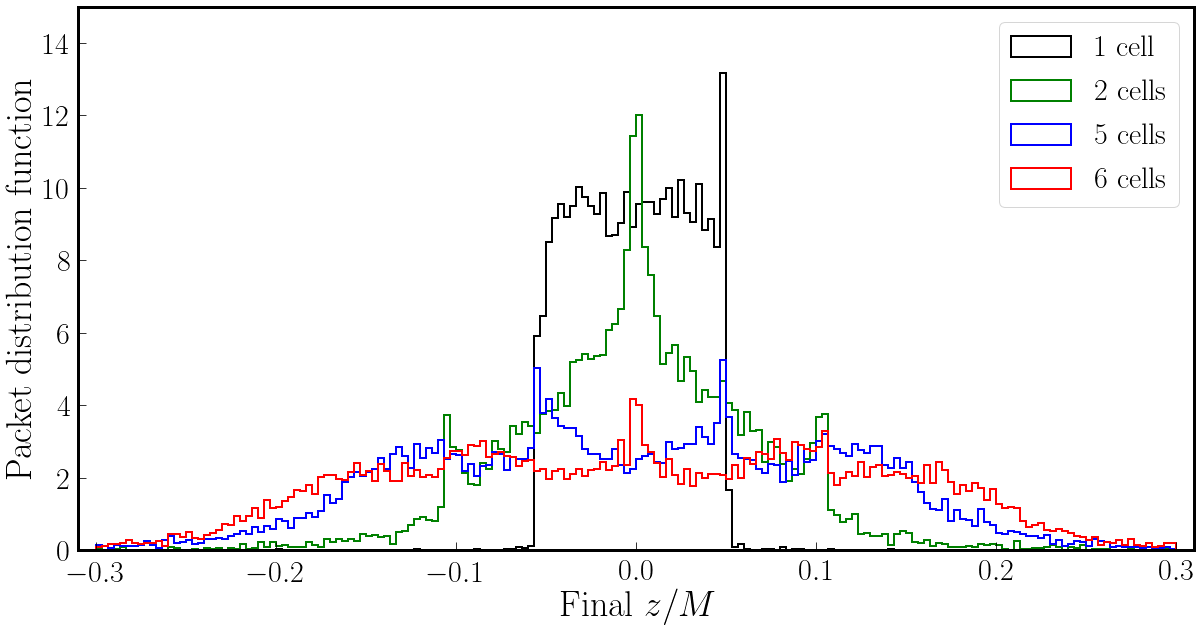}
\caption{Vertical structure of the beam observed in Fig.~\ref{fig:PropagationSchwarzschild}. The initial vertical width of the beam is $[-0.05,0.05]$ (1 cell), $[-0.1,0.1]$ (2 cells), $[-0.25,0.25]$ (5 cells), or $[-0.3,0.3]$ (6 cells). The packets are observed at $x=2.75M$, where the beam should be $\sim 65\%$ of its initial width. Unphysical features in the distribution function observable at cell boundaries decrease in amplitude as the beam becomes more resolved; they do not however entirely disappear.}
\label{fig:PropagationSchwarzschildV}
\end{figure}

We now move to a more direct test of the fact that packets follow the correct trajectory in the Schwarzschild metric. We consider the same computational domain as in our previous low-resolution test, but now emit neutrinos isotropically in the comoving frame of an observer moving with $u_i=(0,100,0)$ within the region $x\in [4,4.6]$, $y\in[0,0.1]$, $z\in [-0.05,0.05]$ (i.e. a region one-cell wide in $y$ and $z$). This practically forces the emission of nearly all packets along the $y$-direction. We then compare the time of arrival and position of the packets to theoretical expectation for the motion of particles along null geodesics. Packets are observed as they cross the $y=2.75$ plane; the results of this test are plotted on Fig.~\ref{fig:PropagationSchwarzschild}. We see that there is very good agreement between theoretical expectations for the trajectory and numerical results. 

Not too surprisingly, our results are not as good when considering the underresolved vertical structure of the beam. Physically, the width of the beam should decrease, with all packets reaching $z=0$ around $x=0$. In practice, this does not happen when the beam is only one-cell thick and we use the value of the metric and of its derivative at the center of a cell everywhere in that cell. Packets within a cell centered on $z=0$ do not experience any vertical acceleration, and the beam thus has constant vertical thickness.  In Fig.~\ref{fig:PropagationSchwarzschildV}, we show the vertical structure of the beam when increasing its vertical thickness from 1 cell to 2, 5, and 6 cells (keeping the cell size constant, and the beam symmetric with respect to the equatorial plane). We see that unphysical subgrid features appear at cell boundaries. These are large for beams 1-2 cells wide, and less prominent for better resolved beams. Using values  of the metric interpolated to the true location of a packet would improve our results, though at a significant computational cost.

Overall, these tests performed in the Schwarzschild metric show that, at the current level of accuracy of numerical simulations, the propagation of packets along null geodesics is unlikely to be a significant source of errors, even with the low-order methods used in this manuscript. 

\subsection{Pair annihilation in crossing beams}
\label{sec:pairtest}

To test our implementation of pair annihilation, we now consider beams of neutrinos crossing at different angles. We use a rectangular grid covering $[-1.2,1.2]\times [-1,1] \times [-0.2,0.2]$, with grid spacing $\Delta x=0.01$ and $G=c=M_\odot=1$. Each beam is emitted from a $0.2\times 0.02 \times 0.01$ region, with neutrinos emitted isotropically in a reference frame with a Lorentz factor $\gamma=1000$ with respect to the simulation frame. This creates a beam of neutrinos narrowly centered along the direction of motion of the emission frame with respect to the simulation frame, but with a nearly uniform distribution of energy in the range $[0,2000\nu_0]$ (for neutrinos of energy $\nu_0$ in the emission frame). We always create that beam in the region $z\in[-0.01,0.01]$, with the $z$ component of the velocity of the emission frame set to $0$. 

We start with a single beam propagating along the $x$-axis. In the emission region, $\eta=10^{-13}$ for $\nu_e, \bar \nu_e$, and $\nu_x$. We find a negligible annihilation rate, as expected for neutrinos and antineutrinos propagating along the same direction. More precisely, the annihilation rate is about six orders of magnitude smaller than when using crossing beams (see below).

\begin{figure}
\centering
\includegraphics[width=0.5\columnwidth]{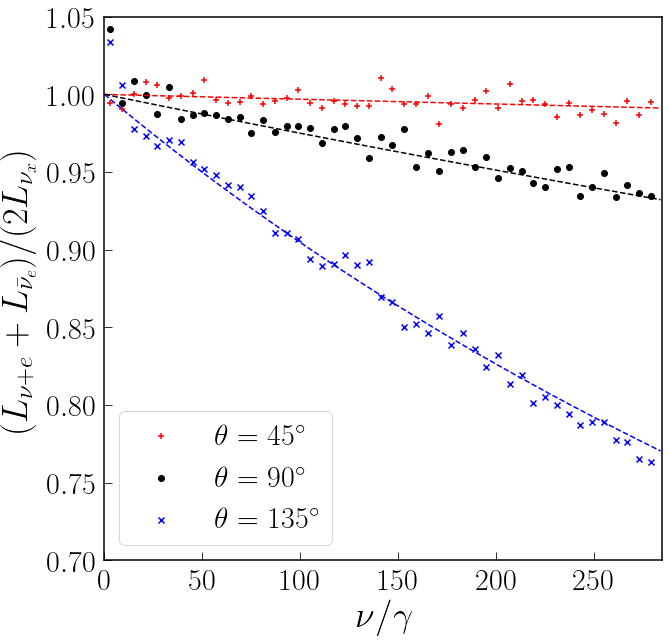}
\caption{Ratio of the average luminosity in $\nu_e$, $\bar \nu_e$ to the luminosity in $\nu_x$ for our crossing beams tests with $\theta=45^\circ$ (red plus symbols),  $\theta=90^\circ$ (black circles), and $\theta=135^\circ$ (blue crosses), plotted as a function of the neutrino energy divided by the Lorentz factor $\gamma=1000$. Dashed lines show theoretical expectations. In these tests, we emit an equal number of neutrinos at $\nu=48\,{\rm MeV}$ and $\nu=142\,{\rm MeV}$ in the emission frame. For each simulation, this plot uses $\sim (7-8)\times 10^6$ Monte-Carlo packets. The noise in the simulation result is due to the random distribution of neutrino momenta in the emission frame, which leads to noise in the energy distribution of neutrinos in the simulation frame.}
\label{fig:pair}
\end{figure}

We then move to two beams with the same emissivity $\eta=10^{-13}$, but crossing at angles $\theta=45^\circ,90^\circ,135^\circ$. As $\theta$ increases, the annihilation rate increases. More specifically, if the energy density of $\bar \nu_e$ in a single beam is $\bar J_e$, then the effective opacity of the region where the two beams interact is, for electron neutrinos of energy $\nu$,
\beq
\kappa_{a,\rm pairs} \approx \frac{C_{\rm pair}cG_F^2}{3\pi} \nu \bar J_e \left(1-\cos\theta\right)^2.
\eeq
If we neglect the changes in $\bar J_e$ due to pair annihilation, and the interaction region has length $L$, we then have
\beq
L_{\nu_e} = L_{\nu_e,0} e^{-\kappa_{a,\rm pairs} L}
\eeq
with $L_{\nu_e},L_{\nu_e,0}$ the luminosity of the beams with and without accounting for pair annihilation. The same result holds for the luminosity of $\bar \nu_e$, while for $\nu_x$ we should replace $\bar J_e$ with $J_x/4$ and use the value of $C_{\rm pair}$ appropriate for heavy-lepton neutrinos. Overall, this leads to an annihilation rate of heavy-lepton neutrinos that is $\sim 20$ times smaller than the annihilation rate of electron-type neutrinos. A more accurate calculation replaces $\bar J_e$ in the computation of $\kappa_{a,\rm pairs}$ by its true value after pair annihilation, which can be done easily be integrating along the direction of propagation of the beams. We use that improved estimate in Fig.~\ref{fig:pair}.

In Fig.~\ref{fig:pair}, we compare theoretical predictions for the neutrino luminosity to our numerical results. We note that our setup allows us to study quite naturally changes in the annihilation rate with the neutrino energy $\nu$, as our relativistic beams produce a wide range of neutrino energies. We find very good agreement between the theoretical predictions and our numerical results.
We note that the theoretical predictions are here derived under the same assumptions as the cross-section used in Sec.~\ref{sec:pairs}, i.e. considering only annihilation of $\nu \bar\nu$ in $e^+ e^-$ pairs, neglecting the mass of the electron, and ignoring blocking factors.

With this setup, we test both the annihilation of neutrinos in the Monte-Carlo code and the way in which we compute the neutrino stress-energy tensor; errors in the stress-energy tensor could easily lead to significant annihilation rates in the single-beam test, or errors in the annihilation rates when using multiple beams.

\subsection{Treatment of high-opacity regions: Spherically symmetric tests}

In Sec.~\ref{sec:highKa} we presented an approximate treatment of regions of high absorption opacities, aimed at limiting the cost of simulations without significantly impacting the energy density and diffusion rate of neutrinos. We can test the error introduced by this modification of the evolution equations on relatively simple spherically symmetric configurations for which we can explicitly evolve the equations of radiation transport. In spherical symmetry, and for neutrinos of a given energy, the distribution function $F(r,\mu)$ only depends on the radius $r$, and the parameter $\mu=\cos\theta$, with $\theta$ the angle between the direction of propagation of the neutrinos. In flat space, Boltzmann's equation is then
\beq
\partial_t F + \mu \partial_r  F + \frac{1-\mu^2}{r} \partial_\mu F = \eta - \kappa_a F + \kappa_s (J-F)
\eeq
with 
\beq
J = \frac{1}{2}\int_{-1}^1d\mu F
\eeq
the energy density. This can easily be written in conservative form if we define $\tilde F = r^2 F$, $\tilde J = r^2 J$, $\tilde \eta = r^2 \eta$:
\beq
\partial_t \tilde F + \partial_r \left(\mu \tilde F\right) + \partial_\mu \left(\frac{1-\mu^2}{r}\tilde F\right) =\tilde \eta + \kappa_s \tilde J - (\kappa_a+\kappa_s)\tilde F.
\eeq
Discretizing this equation on a 2-dimensional grid in $r,\mu$, we can treat the flux terms explicitly using upwind reconstruction, and the source terms implicitly by separately evolving the energy density $\tilde J$
\beq
\partial_t \tilde J + ...= \tilde \eta - \kappa_a \tilde J
\eeq
and the difference between $F$ and $J$ at a given $\mu$, $G=F-J$:
\beq
\partial_t \tilde G +... = - (\kappa_a+\kappa_s) \tilde G.
\eeq
In both cases, the $(...)$ represents the flux terms, which are treated in the explicit part of the time step. This provides us with a convenient test bed for the treatment of high-opacity regions, as we can easily compare results using the correct neutrino-matter source terms $\eta,\kappa_a,\kappa_s$ to results using the approximate values $\eta',\kappa_a',\kappa_s'$ defined in Sec.~\ref{sec:highKa}.

\begin{figure}
\centering
\includegraphics[width=0.45\columnwidth]{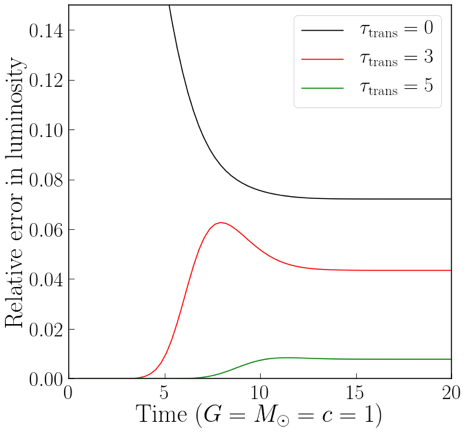}
\includegraphics[width=0.45\columnwidth]{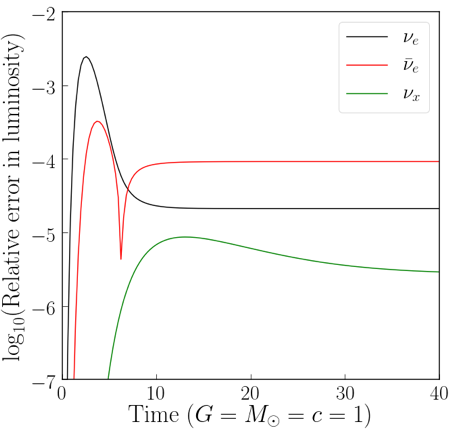}
\caption{Relative error in the neutrino luminosity when using our approximate treatment of high-$\kappa_a$ regions. {\it Left}: Idealized setup where we vary the optical depth $\tau_{\rm trans}$ at which we begin to apply the correction. We see that as $\tau_{\rm trans}$ increases, the error rapidly decreases, down to $\lesssim 1\%$ for $\tau_{\rm trans}=5$. {\it Right}: Realistic profile for $\eta,\kappa_a,\kappa_s$, taken from the polar axis of a binary neutron star simulation, $5\,{\rm ms}$ post-merger~\citep{Foucart:2020qjb}. Errors due to the use of implicit Monte-Carlo techniques are very small in this case ($\ll 1\%$).}
\label{fig:HighKa}
\end{figure}

We start with an idealized configuration: a sphere of radius $r_s=10$, outside of which $\eta=\kappa_a=\kappa_s=0$. Inside, we consider two regions: an outer one with $\kappa_a\Delta r=0.2$ and an inner one with $\kappa_a \Delta r=2$, with $\Delta r = 0.2$ the grid spacing (we also use $\Delta \mu=0.02$). We then evolve the equations of radiation transport with the exact $\eta,\kappa_a,\kappa_s$, and compare to the results of a simulation with $\eta',\kappa_a',\kappa_s'$ corrected in the inner region so that $\eta'/\kappa_a'=\eta/\kappa_a$, $\kappa_a'\Delta r \lesssim 1$, and $\kappa_a+\kappa_s = \kappa_a'+\kappa_s'$ (see Sec.~\ref{sec:highKa}). In Sec.~\ref{sec:highKa}, we argued that the two methods should give similar results if the inner region (where we apply our correction) is well inside of the neutrinosphere. To test this claim, we place the boundary between the inner and outer regions at optical depths $\tau_{\rm trans}=0$ (no outer region), $\tau_{\rm trans}=3$, and $\tau_{\rm trans}=5$. Fig.~\ref{fig:HighKa} shows the relative error in the neutrino luminosity resulting from the use of $\eta',\kappa_a',\kappa_s'$. This figure shows first a transient as the system evolves from $F=0$ to a quasi-equilibrium configuration, and then a time-independent solution at later times. When the corrected region covers the entire emitting sphere ($\tau_{\rm trans}=0$), errors are large: $50\%$ relative errors during the transient (not shown), and $\sim 8\%$ relative errors in steady state. But the error drops quickly as we move the boundary between corrected and uncorrected regions deeper into the emitting sphere. When than boundary is at $\tau_{\rm trans}=5$, we get errors of less than $1\%$. 

In our merger simulations, we have $\tau_{\rm trans}\gtrsim 3$, with even larger values away from the polar regions. In mergers, we also benefit from the presence of a non-zero scattering opacity, and the fact that the opacity inside the neutrinosphere rises much faster than in our simplified test. Both of these effects lead to smaller deviations between the true neutrino energy density and the equilibrium energy density, which should reduce the error induced by our approximate treatment of high opacity regions. To better approximate this error, we consider a spherically symmetric profile for $\eta,\kappa_a,\kappa_s$, with values taken from the polar axis of our BNS simulation using Monte-Carlo transport (using a snapshot at the end of the simulation presented in~\cite{Foucart:2020qjb}). For simplicity, we consider an energy integrated  emissivity, and energy averaged opacities (weighted by the equilibrium energy spectrum of neutrinos). Fig.~\ref{fig:HighKa} shows the resulting relative errors in the neutrino luminosity for all neutrino species. We see that these are much smaller than in our simplified test problem: $\ll 1\%$ for all species during the early-time transient, and $\lesssim 10^{-4}$ in steady state! This provides us with confidence that, for the interactions considered so far (no inelastic scattering), our approximate treatment of high-opacity regions does not significantly impact our numerical results (finite-resolution errors, for example, are much larger than the errors shown on Fig.~\ref{fig:HighKa}). We note that the above results are obtained using $\Delta r=200\,{\rm m}$, to match the resolution of our merger simulation, and $\Delta \mu = 0.02$. The error due to our approximate treatment of high-opacity regions becomes $\sim 2$ orders of magnitude higher when the resolution is decreased to $\Delta r=400\,{\rm m}$, $\Delta\mu=0.04$ (at which point the approximation is used much closer to the neutrinosphere). This confirms the importance of only using our implicit Monte-Carlo prescription well inside the neutrinosphere: if our method is used with overly coarse spatial resolution, or limits $\kappa_a \Delta x$ to values well below $1$, this method will introduce much more significant errors in the evolution.

\subsection{Optically thick sphere in SpEC}

\begin{figure}
\centering
\includegraphics[width=0.7\columnwidth]{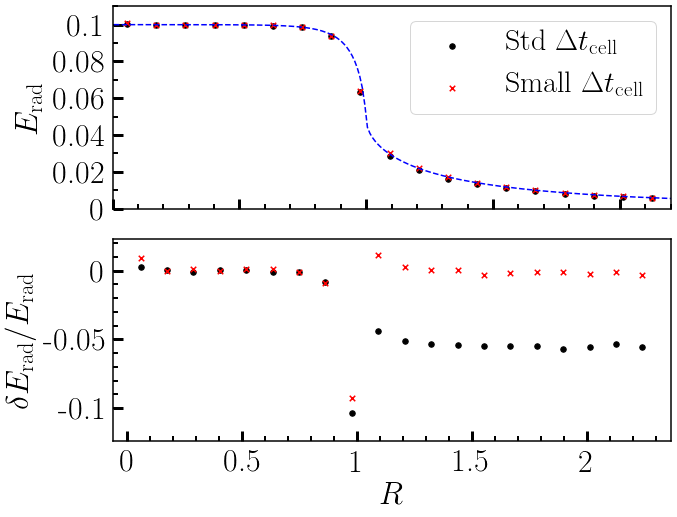}
\caption{ {\it Top}: Energy density of neutrinos in steady-state for a spherically symmetric emitting sphere of radius $r_s=1$ with optical depth $\tau=10$. We show the analytical solution (dashed blue line), as well as results using our standard prescription for the time $\Delta t_{\rm cell}$ that a packet can spend outside of its original cell before we modify the value of $\kappa_{a,s}$ used for its evolution (black circle) and a more accurate simulation using a shorter $\Delta t_{\rm cell}$ (see text). All simulations use $\Delta x=0.025$. {\it Bottom}: Relative error in the energy density of neutrinos for the same test. For the sharp change in $\kappa_a$ used in this test, the choice of $\Delta t_{\rm cell}$ is the dominant source of error in our evolution.}
\label{fig:ka10}
\end{figure}

In the previous section, we discussed the error introduced in the analytical solution of the radiation transport equation when using our approximate treatment of high opacity regions. We now turn to a test allowing us to study the error in finite-resolution simulations, both with and without the use of approximate transport equations. We consider a uniform sphere with fixed emissivity $\eta$ and absorption opacity $\kappa_a$, and with $\kappa_s=0$. The sphere has radius $r_s=1$, and we vary the resolution between $\Delta x=0.05$ and $\Delta x=0.025$. We first consider the case $\kappa_a=10$ (Fig.~\ref{fig:ka10}), for which we always have $\kappa_a \Delta x<1$.  In this case, we do not use our approximations for the treatment of high-opacity region. We also find that the main source of error in the simulations is not the finite grid resolution; instead, it is the value of $\Delta t_{\rm cell}$, the parameter that determines how far a packet can move outside of a given grid cell before we start using the values of $\kappa_{a,s}$ in its new cell. Most packets are moving towards region of lower absorption opacity; thus, if $\Delta t_{\rm cell}$ is large, we effectively overestimate the optical depth of the fluid to neutrinos. This is a particularly large effect in the idealized configuration used here, as the absorption opacity changes from $\kappa_a=10$ to $\kappa_a=0$ instantaneously. We see that for our standard choice of $\Delta t_{\rm cell}$, this introduces a $5\%$ error in our solution. This error is largely independent of the spatial resolution of our simulation (although this is only true because the change in $\kappa_a$ is not smooth). If we instead reduce $\Delta t_{\rm cell}$ to $0.03\Delta x$ (the value typically used in high opacity regions), the relative error in the energy density decreases to less than $0.5\%$. We note that larger errors are observed around $r=r_s$ due to the averaging process used to calculate the neutrino energy density in our simulations.

\begin{figure}
\centering
\includegraphics[width=0.7\columnwidth]{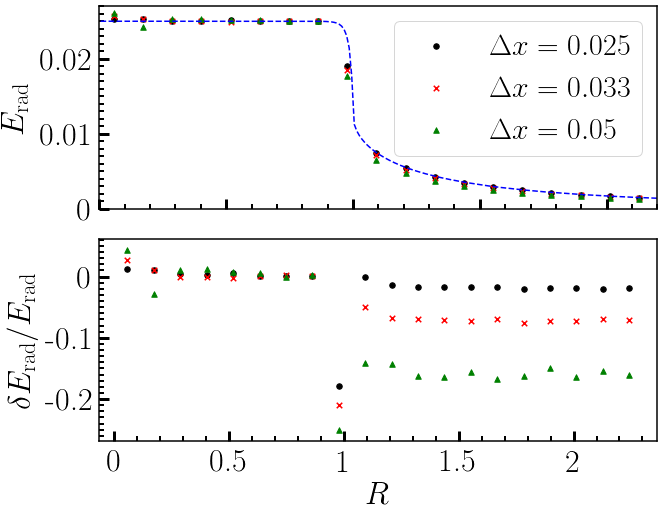}
\caption{Same as Fig.~\ref{fig:ka10}, but for a sphere with optical depth $\tau=40$ and varying the grid spacing (using our standard prescription for $\Delta t_{\rm cell}$). At low resolution $\Delta x=0.05$, each cell has an optical depth $\tau_{\rm cell}=2$, and thus uses our approximate treatment of optical depth region, setting $\kappa_a=\kappa_s=20$ instead of $\kappa_a=40$, $\kappa_s=0$. At high resolution, $\tau_{\rm cell}=1$ and that approximation is not used. For the sharp transition between optically thick and optically thin regions used in this test, the combination of finite $\Delta t_{\rm cell}$, finite resolution, and approximate treatment of absorption lead to $\sim 15\%$ errors at low resolution, i.e. about twice the error in Fig~\ref{fig:HighKa}. The solution does however converge as we increase resolution.}
\label{fig:ka40}
\end{figure}

We then move to a sphere with total opacity $\tau=40$. In this case, a single grid cell has opacity $\tau_{\rm cell}=2$ at our lowest resolution, and $\tau_{\rm cell}=1$ at our highest resolution. As we use an approximate treatment of high-opacity regions when $\tau_{\rm cell}>1$, this probes the transition between using that approximate treatment and evolving the full radiation transport equations. The results are shown on Fig.~\ref{fig:ka40}, for various resolutions $\Delta x$. We see that at low resolution, the combination of finite resolution, approximate treatment of high-opacity regions, and finite $\Delta t_{\rm cell}$ leads to relative errors of $\sim 15\%$. at our highest resolution, the relative error is only $\sim 2\%$. 

An error of $15\%$ would be significant in merger simulations, as differences between Monte-Carlo and two-moment transport are $\sim (10-20)\%$. However, we have shown that this is clearly a worse case scenario for our methods, that we do not expect to encounter in merger simulations. In practice, the opacity will vary more smoothly, significantly reducing the errors due to the finite value of $\Delta t_{\rm cell}$, the finite grid spacing, and the treatment of high-opacity regions (see previous Section).

\subsection{Shock tubes}

One-dimensional shock tubes can also provide us with a relatively simple setup to test our MC code. We consider 3 configurations taken from~\cite{Farris:2008fe}, as adapted for full transport codes by~\cite{Ohsuga_2016}. In each test, we use a grid resolution $\Delta x=0.0125$, with 2 grid points in the $y$ and $z$ directions and periodic boundary conditions along those two axis. We use a time step $\Delta t = 0.25 \Delta x$. We note that while~\cite{Ohsuga_2016} solve these tests in (1+1) dimensions ($x$ and the angle between the $x$-axis and the neutrino momenta), we evolve the full 3D equations, and do not zero the $y$ and $z$ components of the velocity. We use the equation of state $P=\rho T$, $u=P/(\Gamma-1)$, with $P,u$ the pressure and internal energy of the fluid and $\Gamma$ the adiabatic index, which varies between tests. We consider neutrinos in a single energy bin, and only emit $\nu_x$, so that there is no coupling with $Y_e$ (our code is then practically equivalent to a photon transport code). The scattering opacity is $\kappa_s=0$, the absorption opacity $\kappa_a = \rho K$ (with $K$ varying between tests), and the emissivity $\eta = \rho a_{\rm eff} T^4$, with $a_{\rm eff}$ chosen to get a desired equilibrium energy density $J_{\rm eq}=\eta/\kappa_a$ in the fluid frame. We then compare our results to those published in~\cite{Ohsuga_2016} and, as in~\cite{Ohsuga_2016}, verify that in the end state of our simulation the neutrino distribution is a solution to the 1-dimensional, time-independent equations of radiation transport
\beq
\partial_x f(x,\mu) = \frac{\eta \gamma^{-3} (1-v_x \mu)^{-3}-\kappa_a \gamma (1-v_x \mu) f(x,\mu)}{\mu}
\eeq
with $\gamma=\sqrt{1+u_x^2}$ and $v_x=u_x/\gamma$. We solve this equation numerically. 

In our 3D evolutions, $\rho,T,u_i,J$ are initially constant on each side of the $x=0$ plane, with a discontinuity at $x=0$. Monte-Carlo packets at $t=0$ are drawn from an isotropic distribution in the fluid frame, with energy density $J=J_{\rm eq}$. We use frozen boundary conditions for $\rho,T,u_i$. We also redraw the Monte-Carlo packets from an equilibrium distribution function in the fluid frame in cells at the end of our computational domain. We place the outer boundary far enough from the discontinuity in the initial conditions to avoid any effect from reflections at the outer boundary on the regions that we observe. When solving the 1D transport equations, we simply assume $f=f_{\rm eq}$ for the incoming radiation field (i.e. on the left boundary for $\mu>0$ and on the right boundary for $\mu<0$), with
\beq
f_{\rm eq} = \frac{\eta}{\kappa \gamma^4 (1-v_x \mu)^4}.
\eeq

We note that these tests are the only ones in this manuscript to directly test momentum exchanges between neutrinos and the fluid. 

\subsubsection{Non-relativistic strong shock}

\begin{figure}
\centering
\includegraphics[width=0.7\columnwidth]{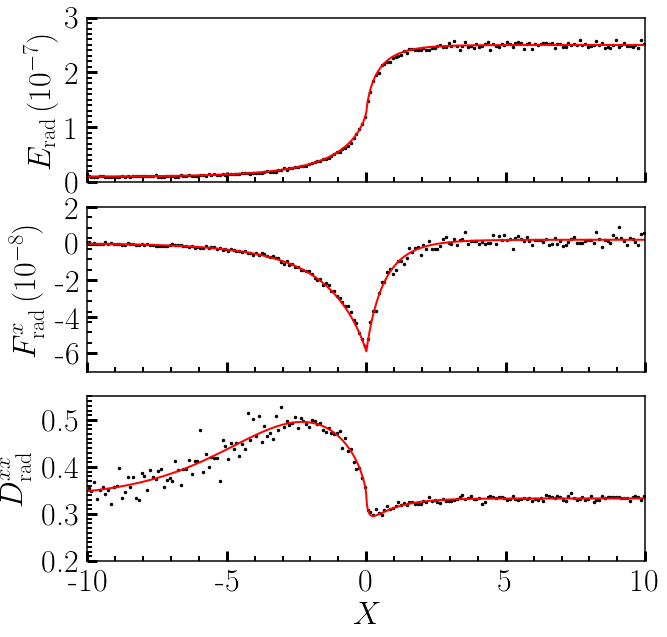}
\caption{{\it Non-relativistic strong shock}: We show the radiation energy density ({\it Top}) and momentum density ({\it Middle}) in the inertial frame, and the Eddington factor ({\it Bottom}) in the fluid frame. Black dots show the results from our Monte-Carlo code, while the solid red line is the solution to the time-independent radiation transport equation for the given fluid variables.}
\label{fig:ST1}
\end{figure}

In this first shock tube test, we consider a stationary shock with a sharp discontinuity and radiation pressure much smaller than the gas pressure. Specifically, for $x<0$ we use $\rho=1$, $P=3\times 10^{-5}$, $u_x=0.015$, $J_{\rm eq}=10^{-8}$, while for $x>0$ we have $\rho=2.4$, $P=1.61\times 10^{-4}$, $u_x=0.00625$, $J_{\rm eq}=2.51\times 10^{-7}$. The adiabatic index is $\Gamma=5/3$ and the specific opacity is $K=0.4$. The radiation pressure is thus $\sim 10^{-3}P$. We use $10^6$ neutrino packets and a domain spanning $x\in [-20,20]$, i.e. $3200\times 2\times 2$ points, and an average of $\sim 80$ packets per grid cell. As we are performing a full 3D evolution, we do not evolve for as long as~\cite{Ohsuga_2016}, but instead stop at time $t=50$. Over that time frame, the shock front remains stationary (i.e. does not move by more than a grid cell). The density shows oscillations at the $\lesssim 20\%$ level within $x\in[-1,1]$, and is otherwise constant. In that same region, the temperature varies at the $\sim 5\%$ level, and $u_x$ at the $2\%$ level, while $|u_{y,z}|\lesssim 10^{-8}$ everywhere. The energy density and flux density of neutrinos in the simulation frame, as well as the Eddington factor in the fluid frame ($D_{xx}=P_{xx}/J$, with $P_{xx}$ the pressure along the $x$-axis in the fluid frame) are shown on Fig.~\ref{fig:ST1}, and compared to the solution of the time-dependent radiation transport equation. In this figure and all other shock tube tests, moments of the neutrino distribution function at a point are calculated using all packets within $\Delta x=0.1$ of that point (i.e. $8\times 2^2=32$ grid cells). Our results are very similar to the full transport solution of~\cite{Ohsuga_2016}, and agree with the steady-state solution within the sampling noise of the simulation. We note that~\cite{Ohsuga_2016} found that in this test the main difference between full transport methods and approximate transport methods was the Eddington factor profile; as in~\cite{Ohsuga_2016}, our Eddington factor is in excellent agreement with the steady-state solution.

\subsubsection{Radiation pressure-dominated shock}

\begin{figure}
\centering
\includegraphics[width=0.7\columnwidth]{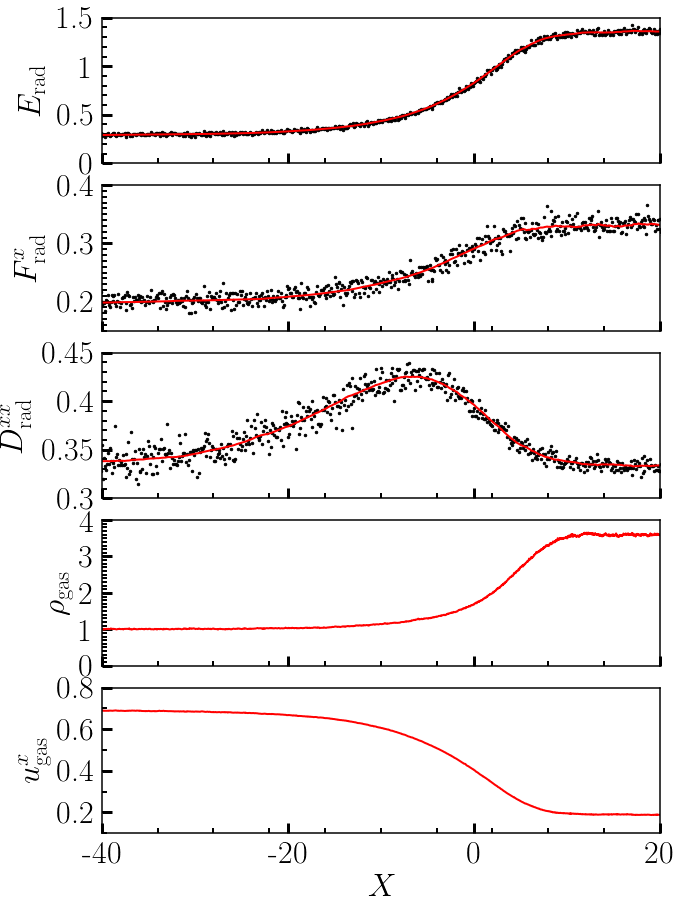}
\caption{{\it Radiation pressure-dominated shock}: Same as Fig.~\ref{fig:ST1}, but for the radiation pressure-dominated shock tube test. We additionally show the result of our coupled radiation-hydrodynamics evolution for the fluid density and $x$-component of the 4-velocity in the bottom two panels; the fluid profile is similar to that found in~\cite{Ohsuga_2016}.}
\label{fig:ST2}
\end{figure}

We now move to a test where neutrino-matter coupling plays a much more important role. We start with $\rho=1$, $P=6\times 10^{-3}$, $u_x=0.69$, $J_{\rm eq}=0.18$ for $x<0$, and $\rho=3.65$, $P=3.59\times 10^{-2}$, $u_x=0.189$, $J_{\rm eq}=1.3$ for $x>0$, with $\Gamma=5/3$ and $K=0.08$. The radiation pressure is thus now $\sim 10P$. We use $6\times 10^6$ neutrino packets and a domain spanning $x\in [-90,90]$, i.e. $14400\times 2\times 2$ points, and an average of $\sim 100$ packets per grid cell.
In this case, the equilibrium shock structure is significantly modified by the presence of radiation. The initial sharp discontinuity in the solution becomes a wider, resolved, radiation pressure-dominated shock. In Fig.~\ref{fig:ST2}, we show the fluid density, fluid velocity, neutrino energy density and neutrino momentum density in the simulation frame, and the Eddington factor in the comoving frame at $t=50$. As in the previous test, we also compare our results with a time-independent solution of the radiation transport equation, using the density, velocity and temperature profiles at $t=200$ in our simulation. We find again good agreement between the MC simulation and this steady-state solution, and between our solution and Fig.13 of~\cite{Ohsuga_2016}. We note that, due to the difference between this steady-state shock solution and our initial conditions, we observe a strong transient over-density traveling in the positive $x$ direction, located at $x\sim 40$ at $t=200$. This transient was not present in the M1 solution of~\cite{Farris:2008fe}, as they initialize their simulation from an exact equilibrium solution of the radiation-hydrodynamics equation; we start instead from an out-of-equilibrium shock that settles into an equilibrium solution. The system nevertheless equilibrates to a steady-state solution with the desired asymptotic values of the fluid density and neutrino distribution function. This test uses an average of $\sim 100$ packets per grid cell, which is similar to the number of particles used in optically thick regions of our merger simulations. As in the previous test, we do not explicitly impose $u_{y,z}=0$, but find that the code maintains $|u_{y,z}|\lesssim 0.01$. 

\subsubsection{Relativistic shock}

\begin{figure}
\centering
\includegraphics[width=0.7\columnwidth]{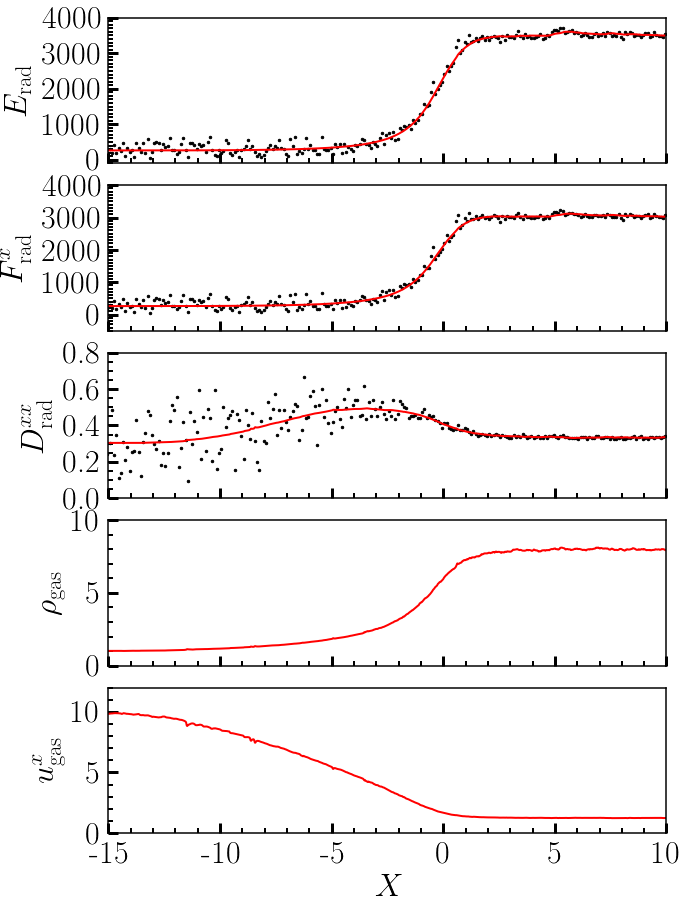}
\caption{{\it Relativistic shock}: Same as Fig.~\ref{fig:ST2}, but for the relativistic shock tube test.}
\label{fig:ST3}
\end{figure}

Finally, we consider a shock tube test with relativistic velocities and temperature, and radiation pressure comparable to the gas pressure. We start with $\rho=1$, $P=60$, $u_x=10$, $J_{\rm eq}=2.0$ for $x<0$, and $\rho=8.0$, $P=2340$, $u_x=1.25$, $J_{\rm eq}=1130$ for $x>0$, with $\Gamma=2$ and $K=0.3$. The fluid is significantly more relativistic than what we typically consider in merger simulations; in fact, in order to perform this test we have to turn off our usual limits on the ratio $T^{\mu\nu}n_\mu n_\nu/\rho_*$ between the energy density of the fluid in the inertial frame and its rest mass density in the inertial frame. We use $10^6$ neutrino packets and a domain spanning $x\in [-40,40]$, i.e. $6400\times 2\times 2$ points, and an average of only $\sim 40$ packets per grid cell. As in the previous test, the solution relaxes from the sharp discontinuity in the initial conditions to a relaxed steady-state with a resolved shock. Fig.~\ref{fig:ST3} shows the state of the simulation at $t=100$, and can be compared to Fig.14 of~\cite{Ohsuga_2016}. This is a very challenging test for our code: the relativistic speeds and high temperatures involved in this test are more extreme that what we typically encounter in the bulk of merger simulations and, more importantly, the radiation is both dynamically important and fairly poorly resolved (as can be seen from the large shot noise in the Eddington tensor, even after aggregating all packets within $32$ neighboring cells!). In this test, we also only maintain $|u_{y,z}|\lesssim 0.15$. The fact that this evolution is stable, recovers the correct average value for the radiation distribution function, and results in profiles for the fluid variables similar to those obtained by~\cite{Ohsuga_2016} indicates that even the low-cost Monte-Carlo simulations used so far in neutron star mergers can handle relativistic shocks with significant radiation pressure without introducing large errors in the simulation.

\subsection{Spherical core-collapse profile}

Finally, we consider a more complex set of tests that allows us to test our implementation of the coupling between the fluid and the neutrinos. This is the main aspects of our algorithm that is expected to be important in merger simulations, yet is not tested by the simpler setups considered so far. We initialize our simulation using a snapshot of the post-bounce remnant of a 3D core-collapse simulation~\citep{Ott2006b}, averaged onto a 1D spherically symmetric profile. We assume that the fluid is at rest, and the metric is Minkowski. We then evolve the equations of radiation transport, accounting for energy transfer and composition changes in the fluid due to neutrino-matter interactions. We do not evolve the equations of hydrodynamics, nor do we deposit linear momentum into the fluid. Practically, this means that the temperature $T$ and electron fraction $Y_e$ evolve due to emission / absorption / scattering of neutrinos, while the fluid density and velocity are kept constant. This setup has been used in a number of tests of neutrino transport algorithms, e.g. in~\cite{Abdikamalov2012,Foucart:2016rxm}. 

We evolve this system at $3$ different resolutions $\Delta x = (6,3,1.5)\,{\rm km}$ while also varying the desired number of packets per species $n_{r,\rm target}=(1,8,64)\times 10^6$ and the desired number of packets in optically thick cells $n_{c,\rm target}=(25,100,400)$. Our grid is a cube extending between $[-300,300]\,{\rm km}$ in each direction, and we evolve the system for $15\,{\rm ms}$. We note that these are fairly low-resolution simulations compared to what is typically used for neutron star mergers. In mergers, we use $\Delta x \lesssim 200\,{\rm m}$, and can thus resolve the neutron star and the region around the neutrinosphere better than in this test. As a result, the region of the simulation covering the proto-neutron star ($r\lesssim 30\,{\rm km}$) is very poorly resolved in this simulation; we are more concerned with the semi-transparent and low-density regions in this test. The value of $n_{c,\rm target}$ and the number of packets present in regions where neutrino-matter interactions are frequent is, on the other hand, comparable to what we used in our published merger simulation using Monte-Carlo transport~\citep{Foucart:2020qjb}. We compare our results with those of the GR1D code.\footnote{https://stellarcollapse.org/gr1d} The GR1D simulation uses a 1D spherically symmetric grid with $\Delta r = 1.5\,{\rm km}$, and is thus more adapted to the geometry of the system than our cartesian grid. 

With these simulations, we can test the convergence of the code. We note however that there is no known analytical solution to this problem, and that GR1D itself uses an approximate two-moment formalism for radiation transport. The GR1D results are however in agreement with the Monte-Carlo results from~\cite{Abdikamalov2012}. What we are testing here is thus consistency between different numerical implementations of the equations of radiation transport as well as the size of finite resolution errors in our simulations, rather than the convergence of the simulations to a known analytical solution to this problem.

\begin{figure}
\centering
\includegraphics[width=0.5\columnwidth]{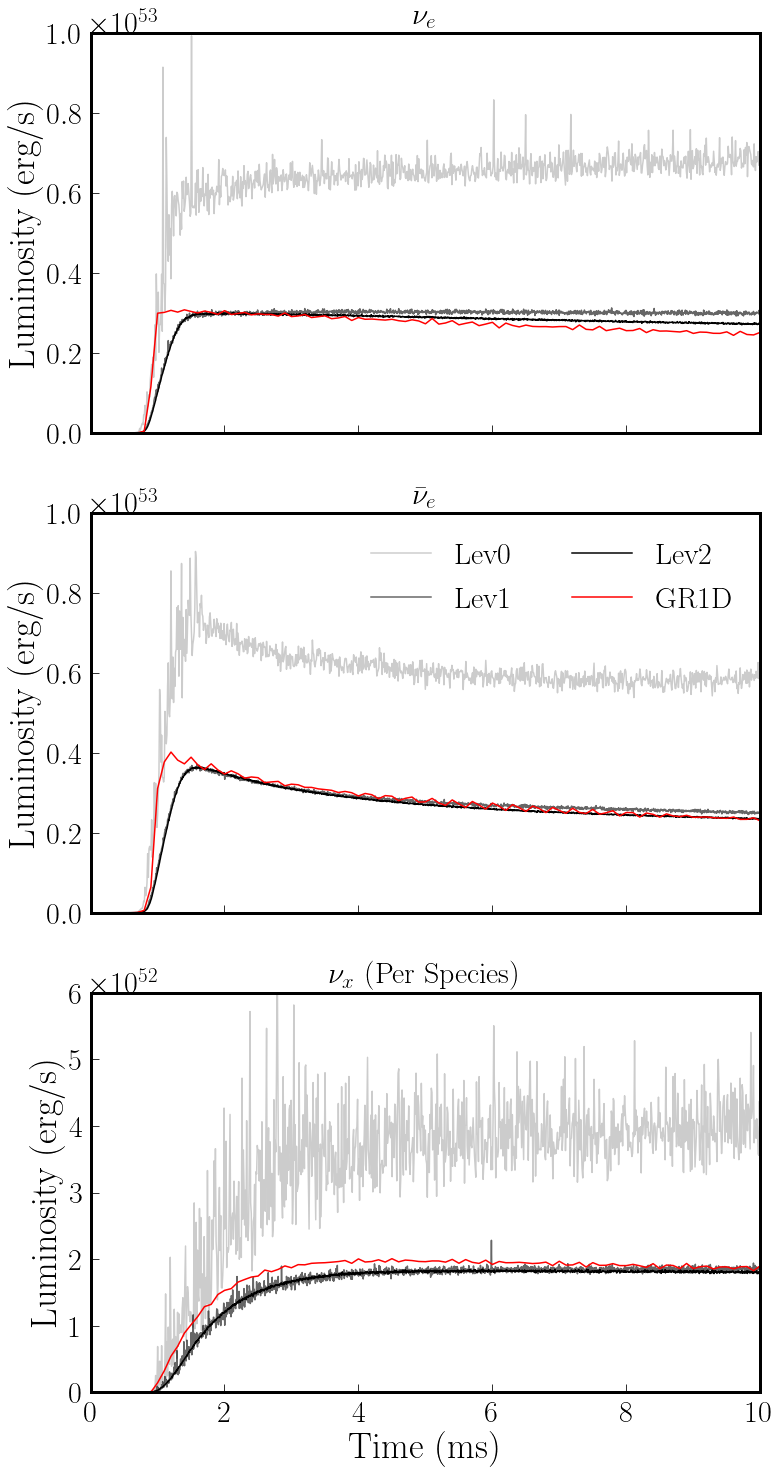}
\caption{Neutrino luminosity for all neutrino species in our simulation of a spherically symmertric core-collapse profile. We show results at $3$ resolutions using our standard computational methods, as well as results from the GR1D code. The difference in the rising timescale of the luminosity is a consequence of the fact that luminosities are measured at the boundary of the computational grids, which is a sphere for GR1D and a cube for SpEC. We find otherwise good convergence of the solution, both in terms of the average value of the luminosity and the amplitude of the shot noise.}
\label{fig:CoreCollapseLum}
\end{figure}

\begin{figure}
\centering
\includegraphics[width=0.5\columnwidth]{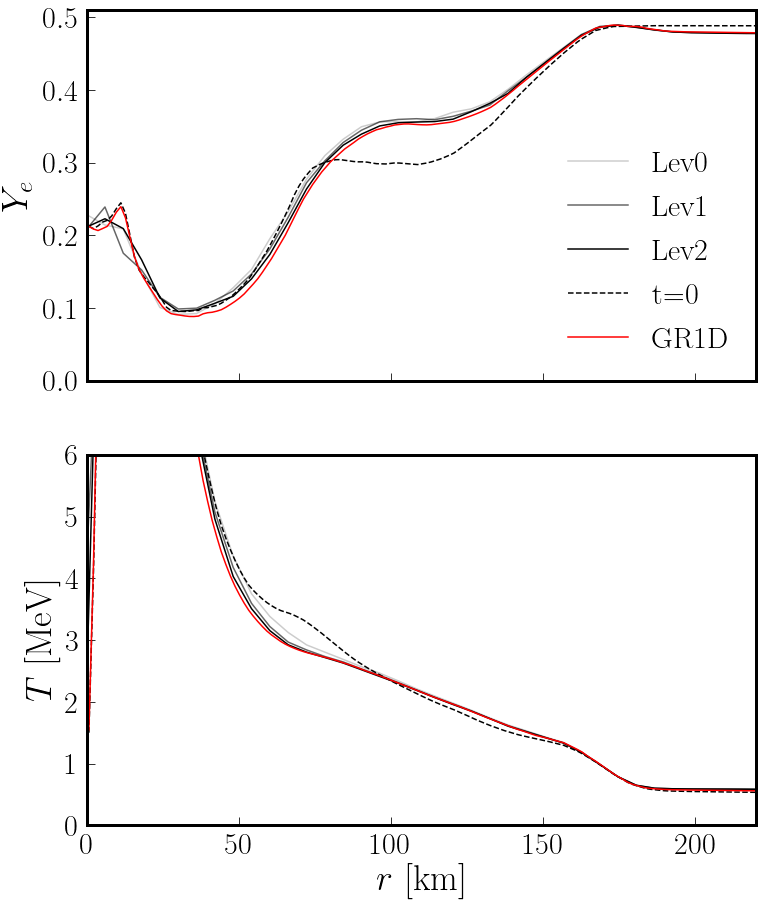}
\caption{Fluid electron fraction ({\it Top}) and temperature ({\it Bottom}) for the same simulations as in Fig.~\ref{fig:CoreCollapseLum}. We also show profiles at $t=0$, for reference. All quantities are measured along the $x$-axis. The neutrinosphere, if calculated by weighting the absorption opacity by the equilibrium energy density of neutrinos, is located at $r=(57,45,21)\,{\rm km}$ for $(\nu_e,\bar \nu_e,\nu_x)$. We enter the region where $\kappa_a \Delta x=1$ at $r=(45,33,15)\,{\rm km}$ in the high-resolution simulation, if we consider the energy-averaged $\kappa_a$. The optically thin and semi-transparent regions are consistent between simulations, while the most optically thick regions are underresolved and inaccurately evolved in this test.}
\label{fig:CoreCollapse8ms}
\end{figure}

Fig.~\ref{fig:CoreCollapseLum} shows the convergence of our Monte-Carlo code with increasing resolution for the neutrino luminosity measured at the boundary of our computational domain. We see convergence of the average value of the luminosity, as well as a decrease in the noise present in the simulations as we increase the number of Monte-Carlo packets. Considering the low resolutions and number of packets used in this test, this is generally encouraging. We should however note one important apparent disadvantage of the Monte-Carlo algorithm: the evolution timescale of the remnant, most easily observable through the decay timescale of the $\nu_e$ luminosity, converges fairly slowly. The approximate two-moment scheme that we used in previous simulations generally showed faster convergence of that evolution timescale at similar resolutions~\citep{FoucartM1:2015}, with even simulations with $\Delta x \sim 6\,{\rm km}$ capturing that timescale as well as the higher resolution Monte-Carlo simulations presented here. At high resolution, we find good agreement with the GR1D results, with differences at early times due mostly to the fact that SpEC measures the neutrino flux at the boundary of a cube, while GR1D measures the neutrino flux at the boundary of a sphere inscribed in that cube.

Fig.~\ref{fig:CoreCollapse8ms} shows the fluid temperature $T$ and electron fraction $Y_e$, $8\,{\rm ms}$ into the evolution. This allows us to test the coupled evolution of the fluid and neutrinos in our simulations. In these profiles, the inner region is a dense neutron star, surrounded by a hot shocked region at $r\sim 20\,{\rm km}$. For $20\,{\rm km}\lesssim r \lesssim 100\,{\rm km}$, neutrino emissions dominate over absorptions, causing the stellar envelope around the remnant to cool.  For $100\,{\rm km}<r<150\,{\rm km}$ neutrino absorptions dominate over neutrino emissions, and the temperature of the fluid increases over time. The electron fraction mostly evolves in the $75\,{\rm km}<r<150\,{\rm km}$, where the fluid becomes progressively more proton rich. We find good agreement between all Monte-Carlo simulations and GR1D on these features. The evolution of the electron fraction is typically better resolved in our Monte-Carlo simulations than in comparable simulations using an approximate two-moment scheme~\citep{FoucartM1:2015,Foucart:2016rxm}, while the heating region is well modeled by all simulations. Looking at the temperature, we find again that the cooling timescale of the Monte-Carlo simulation is convergent, but converges slower than for moment simulations. Inside the proto-neutron star ($r\lesssim 20\,{\rm km}$), our simulations are significantly under-resolved and there is noticeable shot noise in the values of $T, Y_e$ at low resolution. Short timescale variations of the fluid profile in these regions do not strongly impact the evolution of the outer regions or the total luminosity, as their effect is typically smoothed over the (longer) diffusion timescale.

This relatively complex test provides us with an interesting environment to test the impact of many approximations made in our Monte-Carlo scheme. We thus perform another set of simulations, varying numerical methods. We start with a set of simulations identical to our standard case, except that the time step taken by a Monte-Carlo packet is not reduced when that packet gets close to a grid cell boundary. This was the method used for merger simulations in our previous work~\citep{Foucart:2020qjb}. Using this method leads to a mild reduction in the luminosity of $\nu_e$ and $\bar \nu_e$ ($\lesssim 10\%$), and a more significant reduction in the luminosity of heavy-lepton neutrinos ($20\%-40\%$, depending on resolution), most likely due to sharper opacity gradients for heavy-lepton neutrinos. Error cancellations unfortunately led to this method performing better than our improved algorithm at low resolution, especially for the $\nu_x$ luminosity and cooling timescale of the remnant; yet at our two highest resolutions it is clear that it leads to larger errors in the $\nu_x$ luminosity. The evolution of the fluid variables $T,Y_e$ is generally consistent with the results obtained with our current default algorithm. From this test, we conclude that it is possible that part of the difference in $\nu_x$ luminosity observed in~\cite{Foucart:2020qjb} between Monte-Carlo methods and the two-moment formalism (a factor of 2 reduction in luminosity) was due to a suboptimal choice of time step for the evolution of Monte-Carlo packets. On the other hand, we note that Monte-Carlo simulations of a post-merger remnant performed in axisymmetry with a separate code also found higher $\nu_x$ luminosity with Monte-Carlo methods than with a two-moment scheme, indicating that the higher luminosity may very well be physical~\citep{Sumiyoshi:2020bdh}. 

We also consider simulations keeping $\Delta x =3\,{\rm km}$ constant, and varying other aspects of the algorithm. First, we increase by a factor of $2$ the target number of packets per cell and per species ($n_{c,\rm target}$) and the total number of packets per species in our computational domain ($n_{p,\rm target}$). We find a reduction in shot noice, but otherwise no impact on observables. We also vary the minimum scattering optical depth $\kappa_s' \Delta t_c$ beyond which we use our diffusion approximation, increasing it by a factor of $10$. This has no noticeable effect on the results. Finally, we consider changes to the maximum absorption optical depth $\kappa_a' \Delta t_c = \xi$ beyond which some emissions and absorptions are approximately replaced by elastic scatterings, as well as a reduction of the time step used by the Monte-Carlo algorithm (which also results in higher maximum values for $\kappa'_a$). We find that when using $\kappa_a' \Delta t_c > 1$, the accuracy of the solution deteriorates; specifically, the cooling time scale, as measured by the decay rate of the neutrino luminosity, becomes significantly longer. For $\kappa_a' \Delta t_c \gtrsim 2$, the luminosity can even start growing on timescales of $\sim 5-10\,{\rm ms}$. This happens even when the time step is chosen so that $\kappa'_a \Delta t \lesssim 1$. If we increase the spatial resolution at constant $\kappa'_a \Delta t_c$, the solution still improves, but we find best results when $\kappa_a' \Delta t_c \sim 0.5-1$. Practically, it seems that allowing a larger maximum value of $\kappa'_a$ in this test leads to larger discretization errors, and that while these discretization errors converge away as the grid spacing is reduced, they are more significant than errors due to the use of approximate methods in high opacity regions (which is probably not too surprising given the results of the high-opacity tests presented in the previous section). This indicates that even if we could afford to increase $\kappa_a'$ while maintaining the stability of the code, it may not be practically desirable to do so in merger simulations.

\section{Summary}

In this manuscript, we presented the implementation of a cheap Monte-Carlo radiation transport code in the SpEC merger code. This implementation has a cost comparable to that of our two-moment scheme, and to the evolution of the fluid variables themselves~\citep{Foucart:2020qjb}. The main ingredients required to make these evolutions affordable are
\begin{itemize}
\item The ability to distribute Monte-Carlo packets across compute cores rather than tying their evolution to the evolution of the fluid. This is crucial considering that the Monte-Carlo algorithm has very large fluctuations in the number of packets present per cell of the fluid grid.
\item The ability to adaptively choose how many packets are emitted in a given fluid cell, fixing both the total number of packets on the grid and the number of packets sampling the near-equilibrium neutrino distribution in optically thick regions. This allows us to limit both computational costs and sampling noise in the hottest / densest regions.
\item An approximate treatment of scattering events in regions of high scattering opacities, taking advantage of the fact that the evolution of the neutrino distribution function approximately follows a diffusion equation in these regions.
\item An approximate treatment of absorption and emission in regions of high absorption opacities, that effectively limits the mean free path of packets to about one grid cell, without changing the equilibrium energy density and diffusion timescale of neutrino packets. This is the strongest approximation used in our algorithm.
\item The use of low-order methods for the propagation of neutrinos along geodesics, to avoid costly interpolations of metric and fluid variables to the true location of a packet.
\end{itemize}

We improve on the algorithm used in~\cite{Foucart:2020qjb} by allowing Monte-Carlo packets to move from one grid cell to another during a single time step instead of always using the values of the metric, fluid variables, and cross-sections applicable to the cell where they started a time step. We find that this improvement leads to noticeably smaller errors in the diffusion rate of neutrinos in high scattering opacity regions. In merger simulations, this should mostly impact the luminosity of heavy-lepton neutrinos. We also propose a simple method allowing us to account for neutrino-antineutrino pair annihilation in low-density regions without calculating cross-sections for each individual pair of neutrino packets. This new algorithm will allow us to straightforwardly take into account pair annihilation in future merger simulations without having to rely on analytical approximations for the momentum distribution of neutrinos.

We provide both analytical estimates of the errors introduced by these methods and numerical tests of our algorithm. As long as the high absorption opacity regions where we limit the absorption opacity of neutrinos are well inside the neutrinosphere, we expect the approximations proposed in this manuscript to have only a small impact on simulations. The dominant source of errors in simulations is most likely the spatial discretization of the fluid variables. Rapid changes in the fluid density, temperature, and composition indeed lead to rapid changes in neutrino-matter interaction rates and in the equilibrium density of neutrinos. In current merger simulations, where the surface of the post-merger remnant can be poorly resolved (if it is a neutron star) and shocks / turbulence lead to variations of the fluid variables on scales similar to the grid spacing, this is likely to be a more important effect than errors in the propagation, emission, absorption, or scattering of neutrinos.

Monte-Carlo simulations using our relatively cheap implementation should perform very well in semi-transparent and optically thick regions, getting rid of some of the main errors present in moment simulations (energy and momentum closures) while allowing us to robustly implement new physical effects (pair annihilation). In our tests, Monte-Carlo simulations do however have higher errors in the diffusion rate of neutrinos in high opacity regions than moment simulations. This is not particularly surprising, but indicate that Monte-Carlo simulations may become less accurate for long evolutions, when the cooling timescale of the remnant becomes more important than the distribution of neutrinos around it. This trade-off may be possible to avoid with mixed moments/Monte-Carlo methods~\citep{Foucart:2017mbt}, if mixed evolutions can be performed stably, robustly, and cheaply in the future.

\acknowledgments

The authors are grateful to the anonymous reviewer of this paper for helpful comments on an earlier
version of this manuscript, including suggesting many of the tests included in this revised version.
F.F. gratefully acknowledges
support from the NSF through grant PHY-1806278,
from the DOE through grant DE-SC0020435,
and from NASA through grant 80NSSC18K0565.
M.D gratefully acknowledges
support from the NSF through grant PHY-1806207.
H.P. gratefully acknowledges support from the
NSERC Canada. L.K. acknowledges support from NSF grant
PHY-1912081 and OAC-1931280. F.H. and M.S. acknowledge support from NSF Grants
PHY-170212 and PHY-1708213. F.H., L.K. and M.S. also thank
the Sherman Fairchild Foundation for their support. 
Computations were performed on the Plasma cluster at UNH, supported
by the NSF MRI program through grant number AGS 1919310. 
Computations were also performed on the
Wheeler cluster at Caltech, supported by the Sherman
Fairchild Foundation and by Caltech.

\bibliography{References/References}{}
\bibliographystyle{aasjournal}

\begin{table}
\centering
\caption{
  Commonly used symbols and notations (we list here symbols used in more than one section)
}
\label{tab:symbols}
\begin{tabular}{|c|c|}
\hline
Symbols & Interpretation\\
\hline
$x^i, p^\mu$ & Spatial coordinates, 4-momentum \\
$t$ & Time coordinate\\
$\tau$ & Proper time in the reference frame of the fluid\\
$V$ & Coordinate volume of a grid cell\\
$g_{\mu\nu}$ & Spacetime metric\\
$g$ & Determinant of spacetime metric\\
$\alpha, \beta^i, \gamma_{ij}$ & Lapse, Shift, Spatial metric\\
$\rho,T,Y_e, u^\mu$ & Fluid density, temperature, electron fraction, and 4-velocity\\
$T_{\mu\nu,\rm fl}$ & Stress-energy tensor of the fluid\\
\hline
$f(t,x^i,p_i)$ & Neutrino distribution function \\
$\nu_e,\nu_a,\nu_x$ & Electron neutrinos, electron antineutrinos, heavy lepton neutrinos (muon and tau [anti]neutrinos grouped together)\\
$\eta,\kappa_a,\kappa_s$ & Emissivity, Absorption opacity, Scattering opacity as read from NuLib tables\\
$\eta',\kappa_a',\kappa_s'$ & Emissivity, Absorption opacity, Scattering opacity after implicit Monte-Carlo corrections\\
$J,H^\mu$ & Energy density and momentum density of neutrinos in the fluid frame\\
$n_p$ & Total number of Monte-Carlo packets in the simulation\\
$n_{p,\rm target},n_{c,\rm target}$ & Desired number of packets in the simulation ($n_p$) and within an optically thick cell ($n_c$)\\
$N_k$ & Number of neutrinos represented by packet `k'\\
$\nu_k$ & Fluid-frame energy of individual neutrinos in packet 'k'\\
\hline
$\Delta t$ & Full time step in simulation coordinates\\
$\Delta t'$ & Full time step in the fluid frame\\
$\Delta t_c$ & Light-crossing time of a grid cell\\
$\Delta t_{a,s}$ & Time to first absorption/scattering event\\
\hline
\end{tabular}
\end{table}

\end{document}